\documentclass[12pt]{puthesis}

\usepackage{amsmath}
\usepackage{amssymb}
\usepackage{amsfonts}
\usepackage{physics}
\usepackage{url}
\usepackage{lineno}
\usepackage{graphicx}
\usepackage{accents}
\usepackage{bbold}
\usepackage{bm}
\usepackage{longtable}
\usepackage{booktabs}
\usepackage[title,titletoc]{appendix}

\usepackage[T1]{fontenc} % get Ð symbol
\newcommand{\boldD}{\mathord{\textbf{\textit{\DJ}}}} % bold Ð
\newcommand{\italD}{\mathord{\textit\DJ}} % italic Ð
\newcommand{\tens}[1]{\accentset{\leftrightarrow}{#1}} % double over arrow

\usepackage[hidelinks]{hyperref}

\usepackage{cleveref}

\crefname{equation}{Eqn.}{Eqns.}
\Crefname{equation}{equation}{equations}

\usepackage[square, numbers]{natbib}
\author{Brandon Jeremy Bonham}
\adviser{Amitava Bhattacharjee}
\title{The Free-Electron Laser Model of Magnetospheric Chorus}
\submitted{January 2026} % Degree Award Date

\abstract{Chorus waves are electromagnetic waves named for their resemblance to birds chirping at dawn when their radio frequencies are played as audio. The amplification of chorus in Earth's magnetosphere has been the subject of intense scientific inquiry since the discovery of the Van Allen radiation belts in 1958. Resonant interactions between chorus and radiation belt electrons can lead to the exponential growth of small seed waves by a factor of fifty within milliseconds. These powerful modes can cause rapid acceleration of electrons, posing dangers to satellites and other space-based technologies transiting the region. Recent efforts to understand chorus wave amplification have drawn upon parallels to free-electron lasers, laboratory devices that generate intense coherent light with tunable frequencies. This approach, known as the free-electron laser model of magnetospheric chorus, is the subject of this dissertation. In this work, we build on previous research on the free-electron laser model, ultimately presenting a novel nonlinear model of whistler-mode chorus in the magnetosphere. In the first chapter, we provide a brief introduction, including an introduction to whistler waves, magnetospheric chorus, and free-electron lasers. In addition, we derive the 2$N$+2 dynamical equations foundational to the interaction of chorus with $N$ resonant electrons. In the second chapter, we derive a reduced set of just three nonlinear equations for the system using the method of collective variables. From this reduced set, we derive a Ginzburg-Landau equation for the behavior of a chorus wave packet with a spectrum of frequencies with spatially varying amplitudes and discuss the resulting prediction of solitary chorus waves in the magnetosphere. In the third chapter, we focus on the behavior of the single-mode solutions predicted by the GLE, including their linear stability and the phenomenon of mode condensation, where a single mode can emerge from a noisy spectrum of modes with random amplitudes. In the final chapter, we summarize the results and discuss open questions and future directions.}

\acknowledgements{

There are many sayings about the passage of time. People say, for example, that kids grow up faster than you think. They say you blink and a decade is gone. No one has ever said a PhD goes by faster than you think. I would like to acknowledge the colleagues, friends, and family that supported me as a person and a scholar throughout my long journey. 

This starts, of course, with my mother. Mom, I couldn't have accomplished this without the sacrifices you made. Thank you for making me sit at the kitchen table after school every single day until my homework was done, not letting me play the PlayStation or run around outside until I finished. My children will have you to thank for this insightful rule. Thank you for making me feel smart and creating habits that ensured I would be smart. I know now that third grade math really isn't that hard, but thank you for always being so impressed with me. Thank you to my third grade teacher, the late Ms. Samye Smart. You gifted me with a confidence with math that has never left. Dad, thank you for harassing me with multiplication tables, sitting with me for hours in public libraries and half-price bookstores, and teaching me about bits, binary, and Windows 98 all before I was ten. Thank you for talking to me about big ideas. Thank you for making me feel smart. Mom and Dad, thank you for believing in me and always letting me know how proud you are. 

Thank you to my close family, especially my siblings, Ashley, Allison, Dylan and Larissa. You keep my feet on the ground, and make me feel more like myself than anyone. Thank you to my best friend and honorary brother, Ryan. An unconditionally supportive best friend is one of the cheat codes to life. I would not have gotten this far without our long conversations. Thank you to my extended family, including my grandparents, aunts, uncles, cousins, nieces, and nephews. Thank you Lisa, Freddie, and Lettie Cooley. Thank you to my new family, the Phillips family, and in particular Colin and Alyson for treating me like your own. Colin, thank you for teaching me the meaning of the word "banter." Emma, Lori, and Josh, thank you for being an elite hang. 

Thank you to the McNair Scholars Program, without which I would never have gotten my PhD from Princeton. You opened the door for me to compete with those that had much greater advantages. Thank you Rosalinda Luera for greeting me kindly the first time I walked in the office, and every time after that. Thank you Darrell Balderrama and Linda Leech for seeing my potential in the beginning and supporting me through the program. Sonia, I wouldn't have applied to Princeton unless you had asked me, "What's your dream school?" and "Why are you not applying to it?" Thank you Donald Asher for teaching me to "Never self-select out." To my McNair friends, Paola, Ernest, Ashley, Linda, and Parfait, thank you for being my first academic colleagues. Thank you Ronald E. McNair for paving the way for us all. 

The University of Texas at San Antonio was where I decided to be a physicist. Thank you to my dear UTSA friends Lora, Feyi, Korede, Lanston, Hafez, Jason, Adrian, and Herta. Thank you John Cadena for giving me someone to compete with in class. Thank you Anil for being someone I could always discuss math with. Professor Kelly Nash, thank you for giving me someone to look up to. Professor Chonglin Chen, thank you for being so enthusiastic about physics. Professor Koinov, thank you for supervising my first research projects. Professor Guisbiers, thank you for teaching me how to write a scientific paper, and for giving me the opportunity to publish my first one. 

Apart from physics, I have probably learned the most about friendship during this time. Many friendships came late, but were well worth the wait. However, there were some OGs. Thank you Himanshu, Ilya, JP, Christian, Roma, Ho Tat, Maksim, Javier, Ziming, Wayne, Seth, Alex, Lysander, Fedor, Tyler, and Sam. Javier, thank you for inviting me to soccer on my first day at the department. Thank you to the second wave, Jonah, Saumya, Gage, Connor, Gillian, Michael Scheer, Deb, Bendeguz, Zack Gelles, Bennett, Hichem, and Suren. Thank you to my best friends, Sophie, Gio, Adri, Justin, Brad, Adam, and Nick Q. Sophie and Adri, meeting you was a huge turning point for me. If only I had known you sooner. Sophie, thank you for being so enthusiastic and chill about academia. Adri, thank you for being so supportive. Brad, thank you for the camaraderie and the memes. Last, thank you to my fiancée for having such great taste in friends and for letting me have some of them. Finote, Alec, Joel, Kelly, David, Holly, Philipp, Arthur, Qwahn, Harrison, and Emily, you're the coolest and nicest people I know. Holly, I'm so glad we became good friends. If only everyone in the world were Canadian. 

Thank you to the friends who were my greatest teachers. Shouda, thank you for being so helpful to me in discussing mathematics. I have gained much more knowledge from our friendship than you have, but I promise to repay my debt if you ever get interested in physics. Junyi, thank you for being such an exemplary role model. You are still the cleverest graduate student I have ever met. 

There's a special relationship between a black man and his barber. As Patrick the Barber used to say, "If you ain't lookin' good, you ain't feelin' good." Thank you to Mike's Barbershop, including Mike, Oliva, and Bruno for keeping me feeling good all these years. And thank you to Jackie for doing the same for ten dollars cheaper. 

To my Brazilian Jiu Jitsu Club friends, thank you for giving me a home away from the physics department. Why is it so fun to try and beat up your friends? Thank you Jojo, Haley, James, Brie, Jacobo, Bruno, Seyoung, Koda, Lucas, Hinano, Jin, Nick Sudarsky, Eleni, Remy, Amna, and Han. Princeton would not have been the same without you. 

Thank you to Julia Grummitt and Professor Marni Sandweiss for your work on the Princeton \& Slavery Project, which brought to my attention the story of Sam Parker, to whom I have dedicated this thesis. Joseph Henry was an eminent American physicist who made foundational contributions to the study of electricity and magnetism. Sam Parker was a free black man who was provided to Henry by the Trustees to work in his household and laboratory for six years beginning in 1840. Henry referred to Parker interchangeably as "servant" and "assistant," and considered him "indispensable" to the workings of his laboratory. Parker's contributions went largely undocumented and unrecognized in his day and are now mostly lost to history. Interested readers can learn more about Parker, and about other black research assistants such as Alfred Scudder, through the Princeton \& Slavery Project. 

To the incredible university staff, Julio, Darryl, Ted, Angela, Rick, and Anthony, you made the physics department feel more welcoming than anyone. Thank you to the graduate administrators Kate, Katherine, and Leanne. You are the backbone of the graduate program. Leanne, you are way too good at your job. Steph, thank you for always being so friendly and so helpful. Mary Ellen, thank you for your compassion and guidance. 

Thank you to the faculty for all the lessons learned. Professor Pufu, you were the first faculty member I met when I came to Princeton, and left a great first impression of the department. Professor Bialek, thank you for raising the bar as a lecturer. I will forever have something to reach towards. Thank you Kasey Wagoner for being a role model, a mentor, and an ally. Professor Lieb, thank you for emailing me back and encouraging me to keep thinking. Thank you to my first adviser, Professor Jim Olsen. Jim, you were an excellent adviser with a knack for making physics unpretentious. I wish you were an expert in a different subject. We could have done great work together - and I could have graduated years earlier. Thank you Professor Wingreen for being generous with your time when I was searching for a new adviser. Thank you Professor Hantao Ji for teaching me, "There is only one plasma." Thank you to my second reader, Professor Verlinde. The Dutch really are as tall and as discerning as people say. 

Thank you to my committee. You've been so instrumental to my success that I couldn't possibly convey it here. Jo, you have been a role model to me for years. I aspire to be as confident as you. Thank you for looking out for my best interests. Lyman, I wouldn't be here without you. Thank you for seeing it through to the end and being there for me on some of my best and worst days. From you I learned the importance of an unshakable foundation. And I learned to build that mountain "one grain of sand at a time." I learned not to be afraid of hard problems. And I learned how to solve hard problems. I learned how to teach students at the blackboard - by letting them hold the chalk as much as possible. 

Thank you to my adviser. Amitava, you're the youngest person I know. Thank you for welcoming me into your research group. You believed in me so much that you agreed to supervise me for a dissertation in plasma physics before I even knew the definition of a plasma. Your textbook with Professor Gurnett is a work of brilliance. It will always be where I first learned plasma physics. Thank you for teaching me how to do physics at the frontier. Thank you for letting me take full responsibility, and thus full ownership, over my research. You told me once that the skills I learned in solving research problems on my own would give me the confidence to know I could be successful as a physicist. You were right. 

Thank you to the love of my life, Erin. You are the sun. Thank you for never doubting me, and for never letting me doubt myself. I am so much more confident because of you. You taught me that the circumstances I've had to overcome are proof of my strength rather than signs of my weaknesses. You taught me to see the advantages of my differences and that fitting in better with normal people than with physicists doesn't have to be a bad thing. You are my biggest fan and supporter. There are a million more things to say, but I'll save some of them for the wedding. 

}

\dedication{\textit{To Sam Parker \\Research Assistant and Servant of Joseph Henry}}

\begin{document}

%%%%%%%%%%%%%%%
%% Chapter 1 %%
%%%%%%%%%%%%%%%

\chapter{Introduction} \label{Chapter 1: Introduction}

On the evening of Friday, January 31, 1958, the U.S. launched its first satellite, \textit{Explorer} 1 \cite{mcdonald_discovering_2008, koskinen_physics_2022}. Hours later, it was confirmed to be in orbit. The U.S. had officially entered the space race. The craft was equipped with the Iowa Cosmic Ray Instrument, a set of Geiger counters whose live measurements could be transmitted down to Earth. The instrument, built by then graduate student George Ludwig under Professor James Van Allen, was designed to measure cosmic rays in the near-Earth space environment. It performed as expected up to an altitude of about 700 km, when the instrument would fall silent. 

One month later, after a follow-up experiment aboard \textit{Explorer} 2 failed to reach orbit, an identical experiment aboard \textit{Explorer 3} was successfully launched into orbit. This time, an improved Iowa Cosmic Ray Instrument was aboard which included a magnetic tape recorder that could record the data from a complete orbit and relay it back on command to a ground receiving station as it passed overhead. After plotting out the Geiger counter data by hand, noticing the altitude and latitude variations, and the regularity of the pattern, \citet{van_allen_observation_1958} suggested that the instrument was saturated due to the presence of vast regions of energetic particles trapped within Earth's magnetic field. Subsequent missions in the following months, including the \textit{Pioneer} 1 mission which reached an altitude of 17 Earth radii, helped begin to outline the basic structure of an inner and outer radiation belt surrounding Earth. These quickly became known as the Van Allen radiation belts, and were the first major discovery of the space age. 

\begin{figure}
    \centering
    \includegraphics[width=0.7\linewidth]{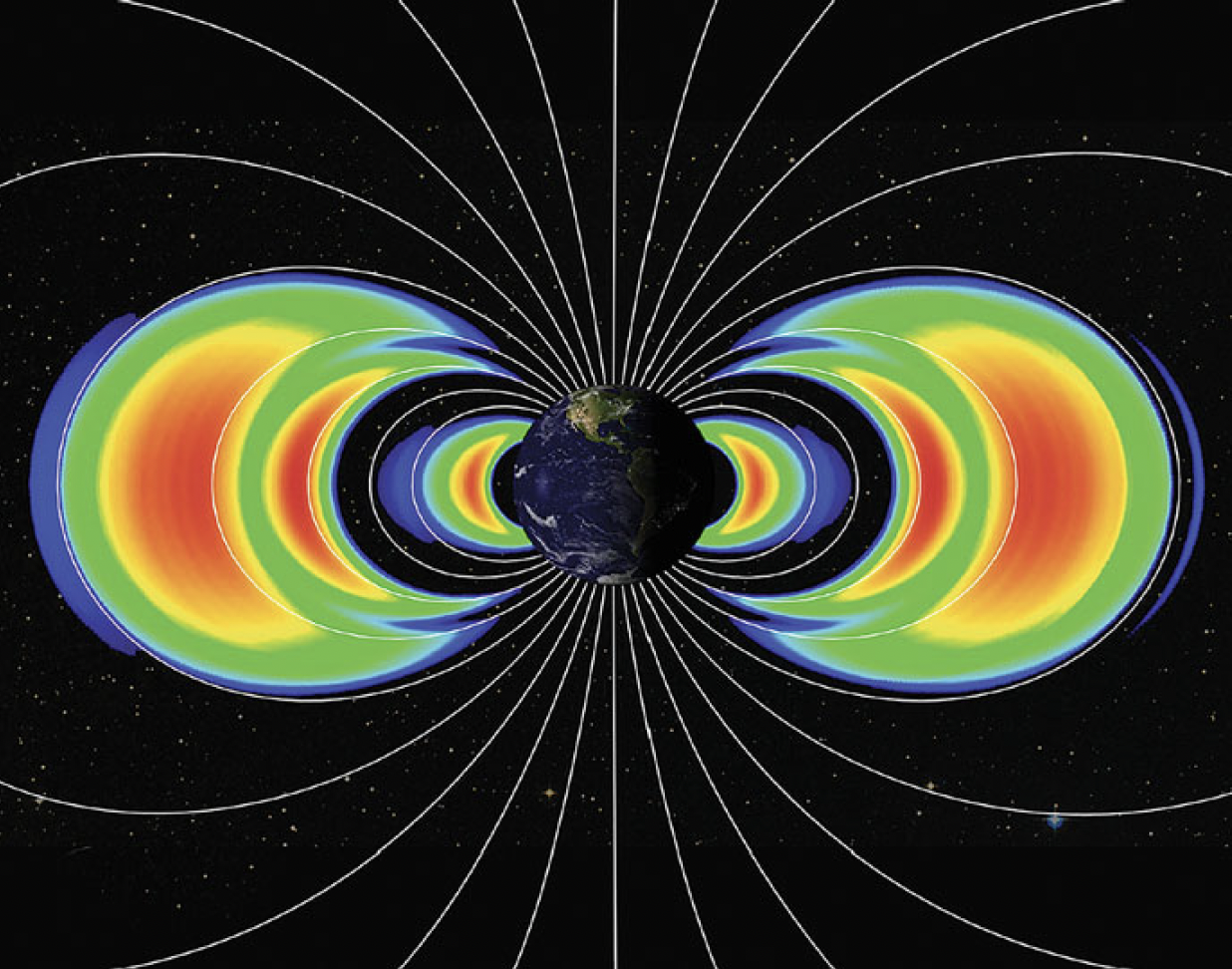}
    \caption{An approximate scale illustration of the basic structure of the inner and outer Van Allen radiation belts. Relative to the center of Earth, the inner belt extends from roughly $1.1-2R_E$ and the outer belt extends from roughly $3-10 R_E$. The magnetic field lines shown illustrate the approximate dipole field of Earth. (Figure reproduced from \citet{koskinen_physics_2022} with original attribution to NASA's Goddard Space Flight Center and The Applied Physics Laboratory of Johns Hopkins University.) }
    \label{Fig: Van Allen Belts}
\end{figure}

The helical motion of the liquid iron alloy deep within the Earth, in its outer core, gives rise to a magnetic field which is approximately a dipole in the main radiation belt domain of $2-7R_E$ \cite{herrmann_geodynamo_2020, koskinen_physics_2022}.\footnote{Altitudes are given with respect to the center of the Earth, and $R_E \approx 6370$ km is the radius of the Earth.} The geometry of a dipole magnetic field can be thought of as a curved magnetic mirror, within which charged particles can become trapped, bouncing back and forth between its mirror points. These trapped particles originate from the upper atmosphere, the solar wind, and cosmic ray sources. The basic structure of the Van Allen belts is illustrated in \cref{Fig: Van Allen Belts}. They consist of two toroidal lobes surrounding Earth concentrated at lower latitudes. The inner belt extends from roughly $1.1-3R_E$ and consists mostly of several MeV to single GeV protons, which corresponds to velocities upwards of ninety-five percent of the speed of light. The outer belt is composed mostly of tens of keV to several MeV electrons, with corresponding velocities upwards of ninety-nine percent of the speed of light, and extends from roughly $3-10R_E$. 

The Van Allen radiation belts and their diverse populations of energetic particles and waves have been the subject of continuous scientific inquiry since their discovery nearly seventy years ago. The interaction of relativistic electrons with resonant electromagnetic waves in the outer, electron-rich belt is the subject of this dissertation. 

\section{Whistler-Mode Chorus}

%% Ch 1 Sec 2.1 %%

\subsection{Introduction} \label{Sec: Ch 1 - Chorus - Introduction}

Due to long-range interactions, plasmas can support a variety of acoustic, electrostatic, and electromagnetic waves which have no counterpart in ordinary states of matter \cite{gurnett_introduction_2017}. There are innumerable types described in the literature, some arise only in specialized circumstances, such as toroidal geometries, while others are nearly ubiquitous in nature. Among other aspects, they can be organized according to their species of origin - ion or electron - and by whether they are due in part to the presence of a background magnetic field. If so, they are further subdivided according to their propagation angle with respect to the background field, which may be parallel, perpendicular, or oblique. The waves with which we are exclusively concerned in this work are the electron species of whistler waves, also known as whistler modes or simply whistlers. In short, a whistler is a right-hand circularly polarized electromagnetic wave which propagates primarily parallel to the background field in a magnetized plasma. 

The discovery of whistlers dates back at least to 1886, when unamplified signals were heard on a 22 kilometer long telephone line in Austria \cite{fuchs1938report}. Similar observations were made independently during an auroral display in 1894 by operators at the British government post office listening in on telephone receivers which were connected to telegraph wires \cite{Preece_1894}. The dispersion relation at the audio frequencies at which whistlers are typically observed dictates that higher frequencies travel faster than lower frequencies. At the observation point, this results in whistling tones that decrease in frequency with time, hence their name. A detailed history of the early scientific investigation of whistlers is given in chapter two of Helliwell's textbook on the subject \cite{helliwell1965whistlers}. 

\begin{figure}[h]
    \centering
    \includegraphics[width=\linewidth]{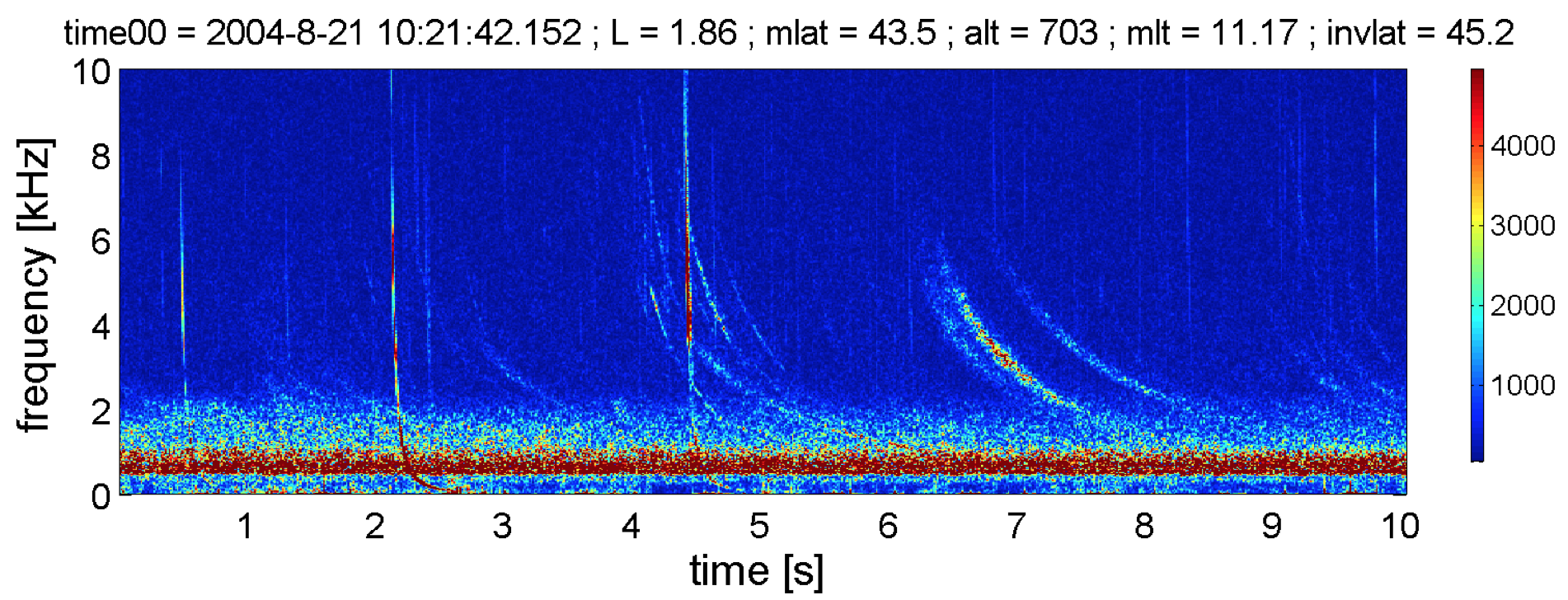}
    \caption{Observation of lightning generated whistler waves in the ionosphere. This electric field spectrogram was created using data captured by the ICE experiment \cite{ice_2006} aboard the DEMETER satellite \cite{DEMETER_2002}. (Figure due to Fiser \textit{et al.} \cite{fiser_whistler_2010})}
    \label{fig: lightning whistlers}
\end{figure}

Although originally observed as an atmospheric phenomenon, whistlers are a general feature of magnetized plasmas, and have been observed in a tremendous variety of circumstances. As a naturally occurring oscillatory mode, they may be excited by the presence of free energy, either internal to the plasma, such as an anisotropic velocity distribution, or external, such as the broad spectrum radiation expelled by lightning discharges. They have been generated in the lab, for example in tokamaks \cite{Heidbrink_2018, spong_first_2018}, and have been induced in the ionosphere by ground-based antennae \cite{helliwell1988vlf}. They have also been observed in a variety of natural environments, including Earth's ionosphere \cite{fiser_whistler_2010} and Van Allen radiation belts \cite{cully2008themis, santolik_spatio-temporal_2003}, and in the solar wind \cite{lacombe2014whistler, kretzschmar2021whistler}. Finally, Voyager 2 observed whistler waves in the magnetosphere of the farthest planet in the solar system \cite{gurnett1990neptunewhistlers}. A typical instance of lightning generated ionospheric whistlers is shown in Figure \ref{fig: lightning whistlers}. 

Planetary radiation belts contain two types of whistler emissions - chorus and hiss - which are primarily generated near the magnetic equator. Hiss consists of a broad, noisy spectrum and plays a major role in radiation belt particle loss \cite{ripoll2017effects, agapitov2020lifetimes}. Chorus is characterized by discrete modes that often sweep rapidly upwards and/or downwards in frequency - a phenomenon known as chirping. Often occurring in quasiperiodic succession, their name is derived from their resemblance to morning birdsong, known as "dawn chorus" in England, when their electromagnetic frequencies are played as audio. This complex frequency-time behavior is related to resonant electron trapping in its rotating wave field. Like whistlers, electrons are also guided by, and rotate in the right-hand sense about the background field. This commonality leads to resonant interactions which can cause rapid energy exchange between the two. Such resonant wave-particle interactions can cause amplification of the waves from a few picoteslas to hundreds of picoteslas and acceleration of the electrons from hundreds of keV to tens of MeV or higher \cite{summers2022analysis, mourenas2023upper}. This high-energy radiation poses risks to satellites and other spacecraft transiting the region, making it essential to understand the amplification mechanisms that govern these waves. 

Early research on whistler waves, roughly during the period between the 1920s and 1950s by Barkhausen \cite{Barkhausen1919}, Eckersley \cite{eckersley1935musical}, Storey \cite{storey1953investigation} and others, focused overwhelmingly on lightning generated whistlers observed by ground-based receivers. Although chorus was among the first whistler emissions observed, it went largely unstudied during this period. A brief history of the theoretical developments subsequent to that period is given in the review article by Hanzelka and Santolík \cite{hanzelka_theories_2024}. The 1958 discovery of the Van Allen radiation belts \cite{van_allen_observation_1958}, two quasi-stable toroidal regions of plasma surrounding Earth, sparked an interest in charged particle acceleration in the magnetosphere. Helliwell and Bell \cite{helliwell1960new}, and others recognized the importance of gyroresonance to the interaction, and their insights were shortly applied to the amplification and damping of the waves, rather than just the acceleration of the particles. In the following decades, research advanced to include crucial concepts foundational to the theoretical analysis of chorus to this day, including electron bunching, whistler phase coherence, and the consistent wave condition. 

%% Ch 1 Sec 2.2 %%

\subsection{Whistler Dispersion Relation} \label{Sec: Dispersion Relation}

One can distinguish and describe plasma waves by their qualitative properties, as done in the previous section, but they are uniquely defined by their dispersion relations. As a proper introduction to the theory of whistlers, in this section we derive their dispersion relation from first principles. This both illustrates how they arise naturally as a linear solution to Maxwell's equations for a cold uniform magnetized plasma, and places them into context as one among a variety of plasma oscillations that can be studied using the same technique. The derivation is standard, but given with particular clarity by Gurnett and Bhattacharjee \cite{gurnett_introduction_2017}. 

The basic problem is to simultaneously solve Maxwell's equations and the Lorentz force equation for the fields and particles which constitute the plasma. We begin with the assumption that the system is primarily in an equilibrium state, but that the particles and fields may execute small oscillations about the equilibrium. Next, we linearize the governing equations under this assumption and rearrange the result into the homogeneous form, $\tens{\boldD} \cdot \bm E = 0$, where $\tens \boldD$ is the desired matrix and $\bm E$ is the electric field.\footnote{Recall that a system of linear equations is homogeneous if all constant terms are zero. When written in matrix form the non-trivial solutions of the system are given by setting the determinant of the matrix equal to zero.} The solution to the homogeneous equation is given by $\det[\tens \boldD]=0$, which ultimately yields the dispersion relation $\italD(\bm k,\omega)=0$. 

First we will sketch the general procedure, applicable to many fundamental plasma waves, then we will apply this procedure to the simplest configuration which exhibits whistler oscillations. We begin by assuming each of the system variables consists of a constant background term and a small sinusoidal perturbation, 
\begin{equation}
\begin{aligned}
    n_s &= n_{s0}+n_{s1}(\bm r,t) \\
    \bm{v}_s &= \bm{v}_{s0} + \bm{v}_{s1}(\bm r,t) \\
    \bm{E} &= \bm{E}_0 + \bm{E}_1(\bm r,t) \\
    \bm{B} &= \bm{B}_0 + \bm{B}_1(\bm r,t),
\end{aligned}
\end{equation}
where $n$ is the number density, $\bm v$ is the velocity, $\bm E$ is the electric field, and $\bm B$ is the magnetic field. The subscript, $s$, denotes the particular species of particle, e.g. electron or ion, and the subscripts zero and one denote the zeroth-order (constant) and first-order (variable) contributions to the variables. The linearization procedure produces sinusoidal solutions, in which case it is easier to work with the Fourier transformed variables. The derivatives in Fourier space are,
\begin{equation}
\begin{aligned}
    \bm \nabla &= i \bm k \\
    \pdv{t} &= - i \omega.
\end{aligned}
\end{equation}

Hereafter we denote the Fourier transformed variables with an overtilde. We can obtain the relation $\tens{\boldD} \cdot \widetilde{\bm E} = 0$ by combining Faraday's and Ampere's laws and eliminating $\bm{\widetilde B}$.  Then the Fourier transform of Faraday's law is,
\begin{equation} \label{Eqn: Faraday Fourier}
\begin{aligned}
    \bm{\nabla} \times \widetilde{\bm E} &= - \pdv{\widetilde{\bm B}}{t} 
    \\
    \Rightarrow i \bm k \times \widetilde{\bm E} &= i \omega \widetilde{\bm B}.
\end{aligned}
\end{equation}
And the Fourier transform of Ampere's law is,
\begin{equation} \label{Eqn: Ampere Fourier}
\begin{aligned}
    \bm{\nabla} \times \widetilde{\bm B} &= \mu_0 \widetilde{\bm J} + \frac{1}{c^2} \pdv{\widetilde{\bm E}}{t} \\
    \Rightarrow i \bm k \times \widetilde{\bm B} &= \mu_0 \tens{\bm{\sigma}}\cdot \widetilde{\bm E} + \frac{1}{c^2} (-i \omega) \widetilde{\bm E}
    \\
    \Rightarrow i \bm k \times \widetilde{\bm B} &= \qty( \mu_0\tens{\bm{\sigma}} - \frac{i\omega}{c^2} \mathbb{1} ) \widetilde{\bm E},
\end{aligned}
\end{equation}
where $\mu_0$ is the permeability of free space, $c$ is the speed of light, $\tens{\bm{\sigma}}$ is the conductivity tensor, $\mathbb{1}$ is the identity matrix, and we have used Ohm's law ($\widetilde{\bm J}=\tens{\bm \sigma} \cdot \widetilde{\bm E}$) to eliminate $\widetilde{\bm J}$. Inserting $\widetilde{\bm B}$ from \cref{Eqn: Faraday Fourier} into \Cref{Eqn: Ampere Fourier} yields the homogeneous equation,
\begin{gather}
    i \bm k \times (\frac{1}{\omega} \bm k \times \widetilde{\bm E} ) = \qty( \mu_0\tens{\bm{\sigma}} - \frac{i\omega}{c^2} ) \widetilde{\bm E} \nonumber
    \\
    \Rightarrow
    \bm k \times (\bm k \times \widetilde{\bm E}) + ( \mu_0 i \omega \tens{\bm{\sigma}} + \frac{\omega^2}{c^2} \mathbb{1}) \widetilde{\bm E} = 0 \label{Eqn: Homogeneous Equation}
    \\
    \Rightarrow \tens{\boldD} \cdot \widetilde{\bm E} = 0. \nonumber
\end{gather}
The conductivity tensor can be determined by equating the Ohm's law definition of the current with particle motion definition
\begin{equation} \label{Eqn: Current Equation}
    \widetilde{\bm J} = \tens{\bm{\sigma}} \cdot \widetilde{\bm E} = \sum_s n_{s} e_s \widetilde{\bm v}_s
\end{equation}
with each $\widetilde{\bm v}_s$ determined from its Fourier transformed Lorentz equation, 
\begin{equation} \label{Eqn: Lorentz Fourier}
\begin{aligned}
    m_s \dv{\widetilde{\bm v}_s}{t} &= e_s (\widetilde{\bm E} + \widetilde{\bm v}_s \times \widetilde{\bm B})
    \\
    \Rightarrow -i \omega m_s \widetilde{\bm v}_s &= e_s (\widetilde{\bm E} + \widetilde{\bm v}_s \times \widetilde{\bm B})
\end{aligned}
\end{equation}

One may work backwards from here to determine the dispersion relation. First determine $\widetilde{\bm v}_s$ from the Lorentz equation (\cref{Eqn: Lorentz Fourier}), then determine $\widetilde{\bm J}$ from the current equation and Ohm's law (\cref{Eqn: Current Equation}), then determine $\tens{\boldD}$ from the homogeneous equation (\cref{Eqn: Homogeneous Equation}), then $\det[\tens{\boldD}] = 0$ yields the dispersion relation, $\italD(\bm k,\omega)=0$. 

Now we employ this procedure to derive the dispersion relation for whistlers by considering the simplest model in which they occur, a cold uniform magnetized plasma. By cold, we mean zero temperature, such that the zeroth-order velocity is zero. This dictates that the zeroth-order electric field is also zero, otherwise the particles would accelerate at zeroth order. Hence, we imagine a plasma with a uniform density $n_{s0}$ in the presence of an externally imposed constant homogeneous magnetic field $\bm B_0$, and determine the linear behavior of the system variables ($n_{s1}, \bm v_{s1}, \bm E_1, \bm B_1$). Hence, 
\begin{equation}
\begin{aligned}
             n_s    &= n_{s0}+n_{s1} \\
    \bm{v}_s &= \bm{v}_{s1} \\
    \bm{E}      &= \bm{E}_1 \\
    \bm{B}     &= \bm{B}_0 + \bm{B}_1.
\end{aligned}
\end{equation}
The linearized Lorentz equation for species $s$ is,
\begin{equation}
    m_s \pdv{\bm{v}_{s1}}{t} = e_s[\bm E_1 + \bm{v}_{s1} \times \bm{B}_0]
\end{equation}
After Fourier transforming the Lorentz force equation, assuming $\bm B_0 \propto \hat z$, and solving for $\widetilde{\bm v}_{s1}$ by matrix inversion, one obtains the current, 
\begin{equation}
    \Rightarrow 
    \left[ \begin{array}{c}
        \widetilde J_x \\
        \widetilde J_y \\
        \widetilde J_z \\ 
    \end{array} \right]
    = \sum_s \frac{n_{s0}e_s^2}{m_s}
    \left[ \begin{array}{ccc}
        \frac{-i\omega}{\Omega^2_{cs}-\omega^2} & \frac{\Omega_{cs}}{\Omega^2_{cs}-\omega^2} & 0 \\
        \frac{-\Omega_{cs}}{\Omega^2_{cs}-\omega^2} & \frac{-i\omega}{\Omega^2_{cs}-\omega^2} & 0 \\
        0 & 0 & \frac{i}{\omega}
    \end{array} \right]
    \left[ \begin{array}{c}
        \widetilde E_x \\
        \widetilde E_y \\  
        \widetilde E_z
    \end{array} \right],
\end{equation}
where $\Omega_{cs}=e_s B_0/m_s$ is the cyclotron frequency for the s\textsuperscript{th} species. Now we may identify $\tens {\bm \sigma}$ from its definition (\cref{Eqn: Current Equation}) and insert it into the homogeneous equation (\cref{Eqn: Homogeneous Equation}) to obtain the explicit form for $\tens{\boldD}$ of $\tens{\boldD} \cdot \widetilde{\bm E} = 0$,
\begin{equation}
    \left[ \begin{array}{ccc}
         S - n^2 \cos^2\theta & -iD &  n^2\sin\theta\cos\theta\\
         iD & S-n^2 & 0\\
         n^2\sin\theta\cos\theta & 0 & P - n^2\sin^2\theta
    \end{array} \right]
    \left[ \begin{array}{c}
         \widetilde E_x  \\
         \widetilde E_y  \\
         \widetilde E_z
    \end{array} \right]
    = 0.
\end{equation}
Hence, 
\begin{equation}
    \tens{\boldD} =
    \left[ \begin{array}{ccc}
         S - n^2 \cos^2\theta & -iD &  n^2\sin\theta\cos\theta\\
         iD & S-n^2 & 0\\
         n^2\sin\theta\cos\theta & 0 & P - n^2\sin^2\theta
    \end{array} \right],
\end{equation}
where, without loss of generality, the coordinate system has been chosen such that the index of refraction (which is parallel to $\bm k$) is in the $x,z$ plane at an angle $\theta$ relative to the background field $\bm n = n(\sin\theta,0,\cos\theta)$. Furthermore, $S$, $D$, and $P$ are functions of $\omega$ given by
\begin{equation}
    S = 1-\sum_s\frac{\omega^2_{ps}}{\omega^2-\Omega^2_{cs}},\ 
    D = \sum_s\frac{\Omega_{cs}\omega^2_{ps}}{\omega(\omega^2-\Omega^2_{cs})},\ 
    P = 1-\sum_s\frac{\omega_{ps}^2}{\omega^2},
\end{equation}
where $\omega^2_{ps} = n_{s0}e^2_s/\epsilon_0m_s$ is the plasma frequency for the $s$\textsuperscript{th} species. Finally, $\det \tens \boldD=0$ for wave propagation parallel to the background field ($\theta=0$) yields the whistler dispersion relation,\footnote{There are two other solutions to $\det \tens \boldD=0$, since $\tens \boldD$ is a $3\times3$ matrix. These correspond to longitudinal electrostatic waves and to the left-circularly polarized waves known as ion cyclotron whistlers.}
\begin{equation}
    n^2 = 1 - \sum_s \frac{\omega^2_{ps}}{\omega(\omega+\Omega_{cs})}.
\end{equation}
In the magnetosphere, typically $n\gg1$, so that the first term may be ignored. Additionally, since whistler frequencies tend to be of order the electron cyclotron frequency, which is roughly a thousand times greater than the proton cyclotron frequency, the electron term dominates the sum. In this case, the dispersion relation can be reduced to the familiar form, 
\begin{equation} \label{Eqn: Cold Whistler Dispersion}
    \frac{c^2 k^2}{\omega^2} \simeq \frac{\omega_{pe}^2}{\omega(\Omega_{e0}-\omega)},
\end{equation}
where $\omega_{pe}$ is the electron plasma frequency and the electron cyclotron frequency, $\Omega_{e0}\equiv e B_0/m_e$, is defined as a positive number. From this expression, we may compute the group velocity,
\begin{equation} \label{Eqn: v_g}
    v_g \equiv \frac{\partial \omega}{\partial k} = 2 c \frac{\omega^{1/2}(\Omega_{e0}-\omega)^{3/2}}{\Omega_{e0}\omega_{pe}}.
\end{equation}
The electric field eigenvector corresponding to this solution is $\widetilde{\bm E}_1 = (E_1,iE_1,0)$, from which Faraday's law gives $\widetilde{\bm B}_1 =(-i B_1, B_1,0)$. One can extract the real fields by inverting the Fourier transform and taking the real part, hence,
\begin{align}\label{Eqn: Whistler Fields}
    \widetilde{\bm E}_1 = \text{Re}[\bm E_1 e^{-i\omega t}] &= E_1(\cos\omega t,\sin\omega t,0) \\
    \widetilde{\bm B}_1 = \text{Re}[\bm B_1 e^{-i\omega t}] &= B_1(\cos(\omega t+\pi/2),\sin(\omega t+\pi/2),0)
\end{align}
where we have assumed $E_1$ and $B_1$ are real. From \Cref{Eqn: Whistler Fields} we can see that the electric and magnetic fields of whistlers rotate in the right-hand sense about the background magnetic field. This simplified picture represents, remarkably well, whistlers observed in a variety of circumstances in nature, including magnetospheric chorus.

%% Ch 1 Sec 3 %%

\section{Free-Electron Lasers} \label{Sec: FEL Introduction}

A free-electron laser (FEL) is a laboratory device that produces intense phase-coherent light with tunable frequencies. While conventional lasers, such as gas and solid-state lasers, generate light at particular frequencies corresponding to discrete atomic or molecular transitions, FELs can produce radiation across a broad and continuous frequency range. While individual devices are themselves tunable, sometimes up to an order of magnitude, the FEL concept in general is so robust that various FEL devices have collectively produced radiation across nearly the entire electromagnetic spectrum, from microwaves to X-rays. In addition, while conventional lasers typically operate at efficiency levels of a few percent, theory indicates that FELs may operate at efficiencies of up to 65\% and efficiencies of up to 40\% have been demonstrated in the laboratory \cite{margaritondo_simplified_2011}. It should be noted that dye lasers are tunable over a narrow frequency range, but operate at comparatively low intensities. 

A brief history of free-electron lasers can be found, for example, in \citet{pellegrini_history_2012} and in the introductory chapter of \citet{freund_principles_2024}. The first source of coherent light from free-streaming electrons was the klystron, invented at Stanford University in the late 1930s \cite{varian1939high}. The device's operation is based on the interaction between radio waves and free electrons propagating within a series of vacuum resonator cavities. Klystron technology was greatly accelerated during World War II, where they played an essential role in radar receivers \cite{caryotakis_klystron_1998}. Undulator-based devices, introduced in 1951 \cite{motz_1951_applications} as a source of coherent emissions at shorter wavelengths than klystrons, were a crucial step towards free-electron lasers. As discussed below, the use of a static undulatory magnetic field forms the basis for free-electron laser operation. In 1960, the ubitron was introduced \cite{phillips_1960_ubitron}. Utilizing free electrons in an undulatory magnetic field and the transverse electric mode in a waveguide, they quickly became the highest power sources of radio-frequency emissions \cite{phillips1988history}. Back at Stanford, four decades after the first coherent light from free-streaming electrons was produced, the FEL concept was proposed \cite{madey1971stimulated}, the term free-electron laser was coined \cite{Madey_FEL_1973}, and the first FEL light was generated \cite{elias1976observation, deacon1977first}. 

For a concise and accessible introduction to the physics of FELs, see \citet{margaritondo_simplified_2011}. More comprehensive treatments can be found in the monographs by \citet{marshall_free-electron_1985}, \citet{brau_free-electron_1990}, \citet{freund_principles_2024}. The basic idea behind FELs is the following. A periodic arrangement of permanent magnets with opposing polarities is used to construct a static sinusoidal magnetic field known as a wiggler or undulator. A beam of relativistic electrons is injected into the wiggler field, causing them to oscillate in the direction transverse to the beam line and emit electromagnetic waves. The Lorentz force from the B-field of the emitted radiation causes the electrons to form bunches and emit increasingly coherent, therefore intense, light.\footnote{The mean intensity of an electromagnetic wave is proportional to the square of the electric field, $\left< I \right> \propto E^2$. If the light is composed of a superposition of $N$ waves with equal amplitudes, $E_0$, and random phases, the individual contributions sum like a random walk, resulting in a combined mean amplitude of $E \simeq E_0 \sqrt N$, hence $\left< I \right> \propto N$. However, if the waves are in phase then the combined mean amplitude is simply $E = E_0 N$, hence $\left< I \right>\propto N^2$.} 

The beam velocity is chosen such that the electrons initially propagate slightly faster than the phase velocity of the radiation field. Just as surfers quickly paddling to catch up with a wave are slowed down as they ascend and ride along with its peak, the electrons slow down as they climb the ponderomotive potential. Hence, the increased intensity of the radiation comes both from the increased phase coherence of the light, and from the transfer of the electron's kinetic energy during the initial bunching process. The intensity saturates (reaches a maximum) when the majority of the electrons become trapped within the troughs of the ponderomotive potential. Thereafter, the electrons continue to oscillate within the potential well, exchanging energy with the radiation field throughout each cycle. This continuous exchange of energy is reflected in post-saturation amplitude oscillations of the radiation field.  

%% Ch 1 Sec 3 %%

\section{Foundational Equations} \label{Sec: Derivation of 2N+2 Equations}

In this section, we present the foundational equations of the FEL model (\crefrange{Eqn: EOM 1}{Eqn: EOM 3}), as derived by \citet{soto-chavez_chorus_2012}, which describe a monochromatic whistler wave interacting with $N$ resonant electrons near the geomagnetic equator. Many authors, dating back decades, have derived and solved similar equations for the particles and fields \cite{omura2011triggering,hanzelka_theories_2024}. However, until the work of \citeauthor{soto-chavez_chorus_2012} they were not cast into the form of FEL equations. The background plasma, although responsible for the presence of the whistler and its dispersion, is assumed to be dynamically unchanged throughout the interaction. It has been shown that under the relevant magnetospheric circumstances, the dominant change in the electron's speed is parallel to the background geomagnetic field, so that the perpendicular speed can be assumed constant throughout, $v_\perp \equiv v _{\perp 0}$. The original equations include an additional term to account for inhomogeneities in the background dipole field as a function of the distance above the geomagnetic equator. However, since chorus amplification is known to take place within a few degrees of this plane \cite{santolik_spatio-temporal_2003}, in this work we focus on the magnetic equator where this term vanishes. 

We begin the derivation with Maxwell's equations, which may be written in potential form as \cite{griffiths2017introduction}
\begin{equation} \label{Eqn: Maxwell's Eqn in Potential Form}
    \nabla^2 \bm A - \mu_0 \epsilon_0 \frac{\partial^2 \bm A}{\partial t^2} - \mu_0\epsilon_0  \frac{\partial (\nabla V)}{\partial t} = -\mu_0 \bm J
\end{equation}
where $\bm A$ and $V$ are the vector and scalar potentials and we have used the Coulomb gauge, $\nabla \cdot \bm A = 0$. We may eliminate the third term on the left side in the following manner. The potential relation
\begin{equation}
    \bm E = -\nabla V - \partial \bm  A/\partial t
\end{equation}
implies the divergence
\begin{equation}
\nabla \cdot \bm E = - \nabla \cdot \nabla V
\end{equation}
where we have again used the gauge condition $\nabla \cdot \bm A = 0$. Combining this with Gauss' law
\begin{equation}
\nabla \cdot \bm E = \rho/\epsilon_0
\end{equation}
yields
\begin{equation}
    -\nabla \cdot \nabla V = \frac{\rho}{\epsilon_0}.
\end{equation}
Differentiating yields
\begin{equation}
    - \nabla \cdot \frac{\partial (\nabla V)}{\partial t} = \frac{1}{\epsilon_0} \frac{\partial \rho}{\partial t}.
\end{equation}
Combining this with the continuity equation
\begin{equation}
    \nabla \cdot \bm J = -\frac{\partial \rho}{\partial t}
\end{equation}
yields
\begin{equation} \label{Eqn: Current Potential Relation}
    \nabla \cdot \frac{\partial (\nabla V)}{\partial t} = \frac{1}{\epsilon_0}\nabla \cdot \bm J.
\end{equation}
By introducing the Helmholtz decomposition
\begin{equation}
    \bm J = \bm J_T + \bm J_L,
\end{equation}
where the "transverse" and "longitudinal" contributions to $\bm J$ satisfy
\begin{align}
    \nabla \cdot \bm J_T  &= 0 \\
    \nabla \times \bm J_L &= 0,
\end{align}
\cref{Eqn: Current Potential Relation} can be written
\begin{equation}
    \nabla \cdot \frac{\partial (\nabla V)}{\partial t} = \frac{1}{\epsilon_0}\nabla \cdot \bm J_L.
\end{equation}
From this we identify
\begin{equation}
    \frac{\partial (\nabla V)}{\partial t} = \frac{1}{\epsilon_0}\bm J_L.
\end{equation}
Then \cref{Eqn: Maxwell's Eqn in Potential Form} can be written,
\begin{equation}
        \nabla^2 \bm A - \mu_0 \epsilon_0 \frac{\partial^2 \bm A}{\partial t^2} - \mu_0 \bm J_L = -\mu_0 (\bm J_L + \bm J_T) 
\end{equation}
or, 
\begin{equation}
        \nabla^2 \bm A - \mu_0 \epsilon_0 \frac{\partial^2 \bm A}{\partial t^2} = -\mu_0 \bm J_T. 
\end{equation}
This is the wave equation from which the dynamical equation for the particles and fields is built. The whistler is assumed to have the form
\begin{equation}
    \bm{B}_w = B_w(t)(\hat{x}\cos{\varphi} + \hat{y}\sin{\varphi})
\end{equation}
with 
\begin{equation} \label{Eqn: cphi def}
    \varphi \equiv \omega(k) t - k z + \phi(t)
\end{equation}
where $\omega(k)$ (hereafter denoted $\omega$) is given by the cold plasma dispersion relation (\Cref{Eqn: Cold Whistler Dispersion}), $k$ is the wave number, $z$ is the direction along the background geomagnetic field, and $\phi(t)$ (hereafter denoted $\phi$) is a time-dependent phase shift. Under a narrow band assumption, which dictates that $B_w$ and $\phi$ are slowly varying functions, the equations for the amplitude and phase may be written as \cite{nunn_self-consistent_1974, omura_review_1991, nunn_numerical_2009, hanzelka_theories_2024}
\begin{gather}
        \frac{d B_w}{dt} = - \frac{\mu_0}{2} v_g J_E 
        \\
        \frac{d \phi}{dt} = - \frac{\mu_0}{2} v_g \frac{J_B}{B_w}.
\end{gather}
where it is assumed that the spatial variation is small compared to the temporal variation, such that the amplitude and phase are only functions of time, i.e. $d/dt = \partial/\partial t + v_g \partial/\partial z \simeq \partial / \partial t$. This is justified insofar as $2\pi/k$ is much less than the distance to the surface of the Earth \cite{soto-chavez_chorus_2012}. $J_E$ and $J_B$ are the resonant electron currents due to the interaction with the whistler, along its electric and magnetic components respectively. They may be written as
\begin{equation}
    \begin{aligned}
        J_E &= e v_{\perp 0} \sum_{j=1}^N \delta(\bm r- \bm r_j) \sin \zeta_j 
        \\
        J_B &= -e v_{\perp 0} \sum_{j=1}^N \delta(\bm r- \bm r_j) \cos \zeta_j
    \end{aligned}
\end{equation}
where $N$ is the number of resonant electrons, $\delta(\bm r -\bm r_j)$ is the Dirac delta distribution as a function of the $j$\textsuperscript{th} electron's position $\bm r_j$, and
\begin{equation} \label{Eqn: zeta def}
    \zeta_j \equiv \theta_j - \varphi
\end{equation}
is the difference between the $j$\textsuperscript{th} electron's gyrophase, $\theta_j$, and the whistler phase, $\varphi$. By defining the complex magnetic field
\begin{equation} \label{Eqn: b def}
    b \equiv \frac{e}{m_e}B_w e^{i \phi}
\end{equation}
and the total complex current
\begin{equation}
    J  \equiv (J_E + iJ_B)e^{i\phi}
\end{equation}
we may combine these into a single equation
\begin{equation} \label{Eqn: Unsimp Complex J and b Equation}
    \frac{d b}{dt} = -\frac{\mu_0 e v_g}{2m_e} J. 
\end{equation}
From the above equations for $J_E$ and $J_B$ we may write
 \begin{equation}
 \begin{aligned}
     J &= e v_{\perp0}\sum_{j=1}^N  \delta(\bm r- \bm r_j) e^{i \phi} \Big( \sin \zeta_j - i \cos \zeta_j \Big)
     \\
     &= -i e v_{\perp0}\sum_{j=1}^N  \delta(\bm r- \bm r_j) e^{i \phi} \Big( \cos \zeta_j + i \sin \zeta_j \Big)
     \\
     &= -ie v_{\perp0} \sum_{j=1}^N \delta(\bm r- \bm r_j) e^{i \phi} e^{i\zeta_j}
     \\
     &= -ie v_{\perp0} \sum_{j=1}^N \delta(\bm r- \bm r_j) e^{i (\phi+\zeta_j)}
     \\
     &= -ie v_{\perp0} \sum_{j=1}^N \delta(\bm r- \bm r_j) e^{i \psi_j}
\end{aligned}
\end{equation}
where the additional $j$\textsuperscript{th} electron phase factor has been defined
\begin{equation} \label{Eqn: psi def}
    \psi_j \equiv \phi + \zeta_j.
\end{equation}
Integrating \cref{Eqn: Unsimp Complex J and b Equation} over the volume
\begin{equation}
        V \frac{d b}{dt} = i\frac{\mu_0 e v_g}{2m_e} \int e v_{\perp0} \sum_{j=1}^N \delta(\bm r- \bm r_j) e^{i \psi_j} d\bm r.
\end{equation}
Thus
\begin{equation} \label{Eqn: Unsimp Complex J and b Equation 2}
    \frac{d b}{dt} = i\frac{\mu_0 e^2 v_g v_{\perp 0}}{2m_eV} \sum_{j=1}^N e^{i \psi_j}.
\end{equation}
One can obtain the expression as written in \citet{soto-chavez_chorus_2012} by employing an $\omega \ll \Omega_{e0}$ approximation for the group velocity. This is justified for lower band whistlers as considered in this work. In this limit the dispersion relation (\cref{Eqn: Cold Whistler Dispersion}) becomes
\begin{equation}
    \frac{c^2 k^2}{\omega^2} \simeq \frac{\omega_{pe}^2}{\omega \Omega_{e0}} 
\end{equation}
so that
\begin{equation}
        \omega \simeq \frac{\Omega_{e0} c^2k^2}{\omega_{pe}^2}
\end{equation}
therefore
\begin{equation} \label{Eqn: v_g appx}
    v_g \simeq \frac{2\Omega_{e0} c^2k}{\omega_{pe}^2}.
\end{equation}
We may also rewrite the dispersion relation  (\cref{Eqn: Cold Whistler Dispersion}) in such a way as to define the parameter $s$,
\begin{equation}
    \frac{c^2 k^2}{\omega_{pe}^2} = \frac{\omega}{ \Omega_{e0}-\omega} \equiv s
\end{equation}
so that \cref{Eqn: v_g appx} may be written as
\begin{equation}
    v_g \simeq \frac{2 \Omega_{e0} s}{k}.
\end{equation}
After employing this form of $v_g$ and the expression for the plasma frequency of the resonant electrons
\begin{equation}
    \omega_{pr}^2 \equiv \frac{e^2 N}{m_e \epsilon_0 V} 
\end{equation}
one may write \cref{Eqn: Unsimp Complex J and b Equation} in the desired form 
\begin{equation} \label{Eqn: bdot equation derived}
    \frac{d b}{dt} = i \frac{s \Omega_{e0} \omega_{pr}^2 v_{\perp0}}{2 k c^2} \frac{1}{N} \sum_{j=1}^N e^{i \psi_j}.
\end{equation}
This expression describes the change in the complex field $b$ due to electron's phase with respect to the whistler. Since $|b| = (e/m_e) B_w$, one can see that the amplitude of the whistler is determined not merely by the electrons' perpendicular velocity, which plays a role in the current which drives the amplitude growth, but also by the electron-whistler phase relationship. 

The Lorentz force equation for the motion of an electron along the Earth's dipole field near the geomagnetic equator
\begin{equation}
    \frac{d (\gamma v_z)}{dt} = |b| v_{\perp0} \sin \zeta
\end{equation}
is well known \cite{dysthe1971some,omura_theory_2008}. This is simpler to solve when written in terms of the proper velocity
\begin{equation}
    \eta_z \equiv \gamma v_z
\end{equation}
where $\gamma = (1-\bm v^2/c^2)^{-1/2}$ is the Lorentz factor. Then, using $\sin\zeta = \tfrac{1}{2i}(e^{i\zeta}-e^{-i\zeta})$, \cref{Eqn: b def}, and \cref{Eqn: psi def} one has
\begin{equation}
    \begin{aligned}
        \frac{d \eta_z }{dt} &= |b| v_{\perp 0 }\frac{1}{2i}\Big(e^{i\zeta}-e^{-i\zeta}\Big)
        \\
        &= |b| v_{\perp 0 }\frac{1}{2i}\Big(e^{i(\psi-\phi)}-e^{-i(\psi-\phi)}\Big)
        \\
        &= \frac{v_{\perp 0}}{2i}\Big( |b| e^{-i\phi}e^{i\psi} - |b| e^{i\phi}e^{-i\psi}\Big)
        \\
        &= \frac{v_{\perp0}}{2} \Big( -ib^*e^{i\psi} +ibe^{-i\psi} \Big).
    \end{aligned}
\end{equation}
Then for each electron, with parallel proper velocity $\eta_z^j$ and phase $\psi_j$, we have the equation of motion
\begin{equation} \label{Eqn: eta dot equation derived}
    \frac{d \eta_z^j }{dt}= \frac{v_{\perp0}}{2} \Big(ibe^{-i\psi_j} +\text{c.c.} \Big).
\end{equation}
Last, we require the equation for the dynamics of $\psi$. Notice from the definitions of $\psi$ (\cref{Eqn: psi def}) , $\zeta$ (\cref{Eqn: zeta def}), and $\varphi$ (\cref{Eqn: cphi def}) that
\begin{equation}
\begin{aligned}
    \psi &= \phi-\zeta \\
    &= \phi + \theta -\varphi \\
    &= \phi + \theta -(\omega t-kz +\phi) \\
    &= \theta - \omega t +kz
\end{aligned}
\end{equation}
therefore
\begin{equation}
\begin{aligned}
    \frac{d\psi}{dt} & = \dot \theta - \omega + k v_z.
\end{aligned}
\end{equation}
The rate of change of the gyro-angle of the electron $\theta$ is given by the relativistic electron cyclotron frequency,
\begin{equation}
    \dot \theta = \frac{e B}{m_e \gamma}.
\end{equation}
Since the whistler amplitude is much smaller than the background geomagnetic field we may approximate $B$ by $B_0$, so that
\begin{equation}
\begin{aligned}
    \dot \theta &= \frac{e B_0}{m_e \gamma} \\
    &= \frac{\Omega_{e0}}{\gamma}.
\end{aligned}
\end{equation}
Then for each electron, with phase $\psi_j$ and Lorentz factor $\gamma_j$, we have the phase equation
\begin{equation} \label{Eqn: psi dot eqn derived}
    \frac{d \psi_j}{d t} =  \frac{\Omega_{e0}}{\gamma_j} - \omega + k\frac{\eta_z^j}{\gamma_j}
\end{equation}
where we have replaced $v_z$ with the proper velocity $\eta_z = \gamma v_z$. 

Together, \cref{Eqn: psi dot eqn derived,Eqn: eta dot equation derived,Eqn: bdot equation derived} constitute the foundational 2N+2 dynamical equations of motion for the whistler-electron interaction as they appear in \citet{soto-chavez_chorus_2012}
\begin{gather}
\frac{d\psi_j}{dt} = \frac{\Omega_{e0}}{\gamma_j} - \omega + k \frac{\eta_z^j}{\gamma_j} \quad\quad (j = 1 \dots N) \label{Eqn: EOM 1}
\\
\frac{d \eta_z^j}{dt} = \frac{v_{\perp 0}}{2} (i b e^{-i \psi_j} + \text{c.c.} ) \quad\quad (j = 1 \dots N) \label{Eqn: EOM 2}
\\
\frac{d b}{dt} = i \frac{s  v_{\perp0}  \Omega_{e0} \omega_{pr}^2}{2kc^2} \frac{1}{N} \sum_{j=1}^N e^{i \psi_j} \label{Eqn: EOM 3}.
\end{gather}
In this form, the one-to-one correspondence with the well-known equations for a single-pass FEL \cite{bonifacio_collective_1986} is manifest,
\begin{gather}
    \frac{d\psi_j}{dt} = \eta_j = \frac{1}{\rho} \left( \frac{\gamma_j}{\gamma_0}-1 \right) \quad\quad (j = 1 \dots N)
    \\
    \frac{d \eta_j}{dt} = -b (e^{\psi_j}+ \text{c.c.}) \quad\quad (j = 1 \dots N)
    \\
    \frac{d b}{dt} = i \delta b + \frac{1}{N} \sum_{j=1}^N e^{-i\psi_j}
\end{gather}
where we have relabeled the variables from the FEL literature to clarify the mapping. The precise interpretation in the FEL context is beyond the aim of this work, however, we note that in both contexts the variables $\psi$, $\eta$, and $b$ assume the role of phase, momentum, and field variables respectively. 

In summary, the dynamics of the whistler-electron interaction has been written in terms of the real variables $\psi$ and $\eta_z$ and the complex variable $b \equiv |b|e^{i\phi}$, which may be determined as functions of time using \crefrange{Eqn: EOM 1}{Eqn: EOM 3}. Physically, the electron-whistler phase difference, $\zeta_j \equiv \theta_j - \varphi$, is related to the variables in the equations of motion through $\zeta_j = \psi_j + \phi$, hence can be determined once $\psi_j$ and $b$ are known; the whistler amplitude is related to $|b|$ through $B_w = (m_e/e)|b|$, hence can be determined once $b$ is known; and the electron parallel velocity is related to the parallel proper velocity through $v_z^j = \eta_z^j/\gamma$, hence can be determined once $\eta_z^j$ is known. One issue is immediately obvious. The behavior of each electron is determined by the behavior of the field, which is itself determined by the behavior of all the electrons, so that even for a moderately sized resonant electron population one must simultaneously solve a large number of coupled differential equations. The reduction of this large number of equations to an analytically tractable set is the first advantage of the free-electron laser model. This can be achieved by the method of collective variables and is the subject of the following chapter. 

%%%%%%%%%%%%%%%
%% Chapter 2 %%
%%%%%%%%%%%%%%%

\chapter{The Free-Electron Laser Model and the Ginzburg-Landau Equation} \label{Chapter 2: FEL and GLE}

%% Ch 2 Sec 1 %%

This chapter is based on work published in \textit{Geophysical Research Letters}, co-authored with Amitava Bhattacharjee \cite{bonham_whistler_2025}. 

\section{Introduction}

As discussed in \cref{Sec: Ch 1 - Chorus - Introduction}, the chorus amplification process has been the subject of theoretical investigation since the 1960s and remains an area of intense research. There have been many successful models, accounting for various aspects of the whistler and electron dynamics, as thoroughly reviewed by \citet{hanzelka_theories_2024}. Recently, it was realized that the dynamics of electrons in the rotating wave field of a whistler wave can be analogized to the dynamics of electrons in the radiation field of a FEL \cite{soto-chavez_chorus_2012}. In this formulation, the role of the wiggler magnetic field in a FEL is played by whistler-mode chorus in the magnetosphere, and the role of the relativistic electron beam in a FEL is played by energetic radiation belt electrons:
\begin{align*}
\text{Wiggler} &\leftrightarrow \text{Whistler} \\
\text{FEL Electron Beam} &\leftrightarrow \text{Radiation Belt Electrons}.
\end{align*}

One may initially pause at this proposition, given that the whistler is a circularly polarized electromagnetic wave and the wiggler is a spatially sinusoidal static plane wave. However, the likeness becomes clear if one notes that a Lorentz transformation along the direction of the beam line renders the wiggler into an electromagnetic wave. \footnote{Recall that a Lorentz transformation from a frame $F$ to a frame $F'$ which is moving at a constant velocity $\bm v$ with respect to $F$ leaves the fields parallel to $\bm v$ unchanged, while the perpendicular fields become,
\begin{align}
    \bm E'_\perp &= \gamma (\bm E_\perp + \bm v \times \bm B_\perp) \\
    \bm B'_\perp &= \gamma (\bm B_\perp - \tfrac{1}{c^2}\bm v \times \bm E_\perp),
\end{align}
where $\gamma = (1-v^2/c^2)^{-1/2}$ and $c$ is the vacuum speed of light \cite{Purcell_Morin_2013}. If in frame $F$ the wiggler field is e.g., $\bm B = B  \cos(kz) \hat x$, then in frame $F'$ the wiggler field is, 
\begin{align}
    \bm B' &= \gamma B \cos[\gamma(k z' - \omega  t')] \hat x \\
    \bm E' &= \gamma v B \cos[\gamma(k z' - \omega  t')] \hat y,
\end{align}
which is the form of an electromagnetic wave. Note, we have also transformed the spacetime coordinates as $z =\gamma (z'+vt')$, and labeled the relative velocity $v = -\omega/k$ so that the wavenumber and frequency of the transformed wiggler are manifest. One may choose $v$ such that the wiggler has the phase velocity of a whistler.} Then the electron beam self-evidently interacts with an electromagnetic wave. Clearer still, since there are only minor differences in the physics of the interaction \cite{pellegrini_history_2012}, we can imagine a helical, rather than a planar wiggler configuration. In which case, in the reference frame of the FEL electron beam, the wiggler field is a circularly polarized electromagnetic wave, and the full analogy to a circularly polarized electromagnetic wave interacting with radiation belt electrons is manifest. 

The analogy between chorus amplification in the magnetosphere and the FEL amplification mechanism in the laboratory was proposed by \citeauthor{soto-chavez_chorus_2012}, who derived \Crefrange{Eqn: EOM 1}{Eqn: EOM 3} for the magnetospheric system which have the form of FEL equations. In this formulation, the role of the wiggler magnetic field in a FEL is played by whistler-mode chorus in the magnetosphere. The whistler is assumed to be produced by another mechanism, such as an anisotropic electron distribution function, but amplified by the FEL mechanism. Using a collective variable approach, \citet{soto-chavez_chorus_2012} derived an analytic expression for the linear growth rate of the wave amplitude and provided estimates for the saturation amplitude and timescale. More recently, the model of \citet{zonca_theoretical_2022}, complemented the FEL analogy by drawing a direct connection between chorus chirping and FEL superradiance. Interestingly, the FEL mechanism has also recently been used to model phenomena in other astrophysical contexts, such as pulsars, magnetars, and fast-radio bursts \cite{fung_free-electron_2004,lyutikov_coherent_2021}. 

In this thesis, we investigate the nonlinear regime of the FEL model of chorus amplification. We employ the model to better understand the nonlinear structure of the whistler wave packets. These packets are often observed to show a high degree of coherence, and we investigate whether these wave packets are solitary wave-like structures (or even more robust structures such as solitons which survive the effect of collisions). These questions are of theoretical as well as observational interest. 

In the original work of \citet{soto-chavez_chorus_2012}, the authors derived 2$N$+2 dynamical equations for the whistler-electron interaction, and reduced these to a set of just three linear equations written in terms of collective variables. Here, we derive the nonlinear collective variable equations, enabling analytical predictions of the saturation amplitude and post-saturation behavior of the wave. We then show that the exponential growth phase and mean saturation behavior can be modeled by the Stuart-Landau equation (SLE) - a simple, yet universal equation for oscillators with a weak nonlinearity \cite{garcia-morales_complex_2012}. In the following chapter, by generalizing to the case of spatial dependence in the amplitude and multiple frequencies, we show that the multi-mode wave amplitude is governed approximately by a Ginzburg-Landau equation (GLE), one of the most celebrated nonlinear equations in physics. 

The GLE has a broad range of applications both within and outside fluid mechanics and plasma physics, having been applied to phenomena such as superfluidity and superconductivity, liquid crystals, and chaotic spirals in slime molds \cite{aranson_world_2002, garcia-morales_complex_2012}. The GLE is known to admit special solitary wave solutions \cite{nozaki_exact_1984} - localized nonlinear wave packets that maintain a definite shape despite being composed of a spectrum of wavelengths with different phase velocities \cite{craik_origins_2004}. They are closely related to solitons, but they lack the elastic scattering property that allows solitons to maintain their shape and speed after colliding with each other. In summary, this work investigates the nonlinear behavior of whistler-mode chorus using the FEL model, strengthens the connection between radiation belt physics and the well-established field of FELs, and provides a framework for the investigation of solitary chorus modes through a novel connection to the GLE. 

%% Ch 2 Sec 2 %%

\section{Theory}

%% Ch 2 Sec 2.1 %%

\subsection{Derivation of Nonlinear Collective Variable Equations} 

Following Bonifacio \textit{et al.} one can define collective variables which can be used to reduce the above 2$N$+2 equations to just three \cite{bonifacio_collective_1986}.  Here, the collective variables are defined as,
\begin{gather}
    A \equiv be^{-i\psi_0} \label{Eqn: AXY Def 1}
    \\
    X \equiv \frac{1}{N} \sum_{j=1}^N e^{i(\psi_j - \psi_0)} \equiv \langle e^{i \Delta\psi}\rangle \label{Eqn: AXY Def 2}
    \\
    Y \equiv \frac{1}{N} \sum_{j=1}^N  e^{i(\psi_j - \psi_0)} (\eta_z^j-\eta_{z0}) \equiv \langle \Delta \eta_z e^{i \Delta\psi}\rangle \label{Eqn: AXY Def 3}
\end{gather}
where $\langle (...) \rangle \equiv \frac{1}{N} \sum_{j=1}^N (...)$ is the average over all electrons, $\eta_{z0} = \gamma_0 v_{z0}$ is the parallel proper velocity of each electron in the initially mono-energetic electron beam, $\gamma_0$ is the initial Lorentz factor, $\Delta \eta_z^j \equiv \eta_z^j - \eta_{z0}$, and $\Delta \psi_j \equiv \psi_j - \psi_0$. Note, the parameter $\psi_0$ is not a constant, but increases linearly with time according to its definition via the detuning constant $\delta \equiv d\psi_0/dt$. This notation is not ideal, but is used in the literature, and is a shorthand for the initial value of the time derivative of $\psi$, $\left. d\psi/dt \right|_{t=0} \equiv d\psi_0/dt$. Therefore, $\delta \equiv d\psi_0/dt = \Omega_{e0}/\gamma_0 - \omega + k \eta_{z0}/\gamma_0$, or $\psi_0 = \delta t+(\text{const.})$, where the constant is here assumed to be zero. The detuning parameter can be written more instructively as $\delta=k(v_{z0}-v_r)$, where  $v_r\equiv(\omega-\Omega_{e0}/\gamma)/k$ is the electron resonance velocity. 

The collective variables, $A$, $X$, and $Y$, can be thought of as amplitude, electron phase, and electron momentum variables respectively. In particular, $|A|=(e/m_e)B_w$ is proportional to the amplitude of the whistler. Or with the complex phase, it can be used to determine the phase of the whistler, $A = (e/m_e)B_w e^{i(\phi-\delta t)}$, so that $\phi = \varphi_A +\delta t$, where $\varphi_A$ is the complex phase of $A$, i.e. $A = |A|e^{i\varphi_A}$ . The variable $X$, known in the FEL literature as the bunching variable, measures the degree of randomness of the electron phases. As can be seen from the definition, if the electron phases are random, then $X$ tends to zero, and if the electrons all have the same phase, then $|X|=1$. The variable $Y$,  being a mixed variable has a less direct interpretation, but plays the role of a phase weighted momentum. 

To obtain the collective variable form of \Crefrange{Eqn: EOM 1}{Eqn: EOM 3} we take time derivatives of the collective variable definitions, \Crefrange{Eqn: AXY Def 1}{Eqn: AXY Def 3}, and replace terms containing $\dot{\psi}, \dot{\eta_z}$, and $\dot b$ with their definitions from the original equations of motion, \eqref{Eqn: EOM 1} - \eqref{Eqn: EOM 3}. 

First we derive $A$, which is an exact relation of the equations of motion. From \cref{Eqn: AXY Def 1}, 
\begin{equation}
\begin{aligned}
    \dot A &= \frac{d}{dt} \Big( b e^{-i\psi_0}\Big) \\
    & = \dot b e^{-i \psi_0} - ib\dot\psi_0 e^{-i \psi_0}.
\end{aligned}
\end{equation}
We can simplify the first term by using \cref{Eqn: EOM 3} with $g \equiv \frac{s v_{\perp0}\Omega_{e0}\omega_{pr}^2}{2 k c^2}$, and we can simplify the second term using the definition $\dot \psi_0 = \delta$. Thus, 
\begin{equation}
\begin{aligned}
    \dot A &= \Big( i g\sum_{j=1}^N e^{i \psi_j} \Big) e^{- i \psi_0} - i \delta b e^{-i\psi_0} \\
    &= i g \sum_{j=1}^N e^{i(\psi_j - \psi_0)} - i \delta A \\
    &= i g X - i \delta A
\end{aligned}
\end{equation}
The collective variable equations for $X$ and $Y$ are also obtained by taking time derivatives of their definitions, but additionally make use of expansions of $\gamma$ and $1/\gamma$ about the initial proper velocity,
\begin{align}
    \gamma &= \gamma_0 + \Gamma_1 \Delta \eta_z + (1/2)\Gamma_2\Delta \eta_z^2 + O(\Delta \eta_z^3) \\
    \gamma^{-1} &= \gamma_0^{-1} + \Gamma_{\scriptscriptstyle -1} \Delta \eta_z + (1/2)\Gamma_{\scriptscriptstyle -2}\Delta \eta_z^2 + O(\Delta \eta_z^3),
\end{align}
where for convenience we have written the $n$\textsuperscript{th} derivatives of $\gamma$ and $1/\gamma$ as,
\begin{align}
    \Gamma_n &= \left. \frac{d^n\gamma}{d\eta_z^n} \right |_{\eta_{z0}} \\
    \Gamma_{-n} &= \left. \frac{d^n\gamma^{-1}}{d\eta_z^n} \right |_{\eta_{z0}} .
\end{align}
Since $v_{\perp}=v_{\perp0}$ is a constant the form $\gamma = \gamma_{\perp0} (1+\eta_z^2/c^2)^{1/2}$, where $\gamma_{\perp0} \equiv (1-v_{\perp0}^2/c^2)^{-1/2}$, may be used to most easily compute the derivatives.\footnote{The Lorentz factor may be written in terms of $\gamma_{\perp0} \equiv (1-v_{\perp0}^2/c^2)^{-1/2}$ via,
\begin{equation}
\begin{aligned}
    \gamma &= (1+\eta_\perp^2 / c^2 + \eta_z^2/c^2)^{1/2} \\
    &= (1+ \gamma^2v_{\perp0}^2/c^2+\eta_z^2/c^2)^{1/2} \\
    \Rightarrow \gamma^2 &= 1 + \gamma^2 v_{\perp0}^2/c^2 + \eta_z^2/c^2 \\
    \Rightarrow \gamma^2(1-v_{\perp0}^2/c^2) &= 1 + \eta_z^2/c^2 \\
    \Rightarrow \gamma &= \gamma_{\perp0}(1 + \eta_z^2/c^2)^{1/2}
\end{aligned}
\end{equation}
} 

Thus,
\begin{equation}
\begin{aligned}
    \Gamma_1 &= \frac{\gamma_{\perp0}^2 \eta_{z0}}{\gamma_0 c^2}\\
    \Gamma_2 &=  \frac{\gamma_{\perp0}^4}{\gamma_0^3 c^2} \\
    \Gamma_{-1} &= -\frac{\Gamma_1}{\gamma_0^2} = -\frac{\gamma_{\perp0}^2\eta_{z0}}{\gamma_0^3 c^2}\\
    \Gamma_{-2} &= \frac{2 \Gamma_1^2}{\gamma_0^3} - \frac{\Gamma_2}{\gamma_0^2} = \frac{\gamma_{\perp0}^4}{\gamma_0^5 c^2}\Big( \frac{2 \eta_{z0}^2}{c^2} -1 \Big)
\end{aligned}
\end{equation}
Now we may obtain the collective variable equation for $X$. Taking the derivative of \cref{Eqn: AXY Def 2} and using \cref{Eqn: EOM 1}, 
\begin{equation}
    \begin{aligned}
        \dot X &= \left< i\dot{\Delta \psi} e^{i \Delta \psi}\right> \\
        &= i \left< (\dot{\psi} - \dot\psi_0) e^{i \Delta \psi} \right> \\
        &= i \left< e^{i\Delta \psi}\left[ \Big(\frac{\Omega_{e0}}{\gamma} - \omega + k \frac{\eta_z}{\gamma}\Big) - \Big(\frac{\Omega_{e0}}{\gamma_0} - \omega + k \frac{\eta_{z0}}{\gamma_0}\Big) \right] \right> \\
        &= i \left< e^{i\Delta \psi} \left[ \Omega_{e0}\Big( \frac{1}{\gamma} -\frac{1}{\gamma_0} \Big) + k \Big(\frac{\eta_z}{\gamma} -\frac{\eta_{z0}}{\gamma_0}  \Big) \right] \right>. 
    \end{aligned}
\end{equation}
Expanding $1/\gamma$ to $O(\Delta\eta_z^2)$, 
\begin{equation}
    \begin{aligned}
        \dot X &\simeq i \bigg< e^{i\Delta\psi} \Big[ \Omega_{e0}\Big( \gamma_0^{-1} + \Gamma_{-1} \Delta\eta_z +\tfrac{1}{2} \Gamma_{-2} \Delta\eta_z^2 - \gamma_0^{-1} \Big) 
        \\
        & \qquad \qquad + k \eta_z \Big( \gamma_0^{-1} + \Gamma_{-1} \Delta\eta_z +\tfrac{1}{2} \Gamma_{-2}\Delta\eta_z^2 \Big) - k \eta_{z0} \gamma_0^{-1}\Big] \bigg> \\
        &\simeq i\left <  e^{i\Delta\psi} \Big[ \Delta\eta_z\Big( \Gamma_{-1} \Omega_{e0} + k\gamma_0^{-1} + k \Gamma_{-1} \eta_{z0} \Big) 
        + \Delta\eta_z^2 \Big( \tfrac{1}{2} \Omega_{e0}\Gamma_{-2} + k \Gamma_{-1} + \tfrac{1}{2} k \eta_{z0} \Gamma_{-2} \Big)\Big] \right> \\
        & \simeq - i h_c \left < \Delta \eta_z e^{i \Delta \psi}\right> + i \zeta_0 \left< \Delta \eta_z^2 e^{i \Delta \psi} \right> \\
        & \simeq - i h_c Y + i \zeta_0 \left< \Delta \eta_z^2 e^{i \Delta \psi} \right>, 
    \end{aligned}
\end{equation}
where we have used the relation $\eta_z = \Delta\eta_z +\eta_{z0}$, and the constants $h_c$ and $\zeta_0$ are defined by, 
\begin{align}
    h_c &\equiv -(\Gamma_{-1} \Omega_{e0} + k \gamma_0^{-1} + k \Gamma_{-1} \eta_{z0}) 
    = -  \frac{k}{\gamma_0}\Big( 1 - \frac{\gamma_{\perp0}^2{\eta_{z0}^2}}{\gamma_0^2 c^2} - \frac{\gamma_{\perp0}^2 \Omega_{e0}\eta_{z0}}{\gamma_0^2c^2k} \Big) \label{Eqn: h_c} \\
    \zeta_0 &\equiv \tfrac{1}{2} \Omega_{e0} \Gamma_{-2} + k\Gamma_{-1} + \tfrac{1}{2}  k \eta_{z0}\Gamma_{-2}. 
\end{align}
Next we obtain the collective variable equation for $Y$. Taking the derivative of \ref{Eqn: AXY Def 3}, 
\begin{equation} \label{Eqn: Y intermediate}
    \begin{aligned}
        \dot Y &= \left< \dot{\Delta \eta_z}e^{i\Delta\psi} + i\Delta\eta_z \dot{\Delta\psi} e^{i\Delta\psi} \right>.
    \end{aligned}
\end{equation}
We will simplify this term by term. Using \cref{Eqn: EOM 2} and \cref{Eqn: AXY Def 1}, the first term may be written, 
\begin{equation}
\begin{aligned}
    \left< \dot{\Delta\eta_z}e^{i\Delta\psi} \right> &= 
    \left< \left( \dot \eta_z-\dot\eta_{z0} \right)e^{i\Delta\psi} \right> \\
    &= \left< \dot \eta_z e^{i\Delta\psi} \right> \\
    &= \left< \Big(i u b e^{-i\psi} - i u b^*e^{i\psi}\Big) e^{i(\psi-\psi_0)} \right> \\
    &= \left< i u b e^{- i \psi_0} - i u b^* e^{i(2\Delta\psi+\psi_0)} \right> \\
    &= i u A - i u A^* \left< e^{2i\Delta\psi} \right> \\
    &\simeq i u A. 
\end{aligned}
\end{equation}
Where the term $-iuA^*\langle e^{2 i \Delta\psi} \rangle$ may be dropped since it has been shown to be negligible contribution to the resulting nonlinear equations \cite{bonifacio_collective_1986}. The second term in \cref{Eqn: Y intermediate} may be simplified by the same procedure as was employed for $\dot X$, 
\begin{equation}
    \begin{aligned}
        \left< i \Delta\eta_z \dot{\Delta\psi} e^{i \Delta \psi} \right> &= i \left< \Delta\eta_z (\dot \psi-\dot\psi_0)e^{i\Delta\psi} \right> 
        \\
        &= i \left< \Delta\eta_z e^{i\Delta \psi}\left[ \Big(\frac{\Omega_{e0}}{\gamma} - \omega + k \frac{\eta_z}{\gamma}\Big) - \Big(\frac{\Omega_{e0}}{\gamma_0} - \omega + k \frac{\eta_{z0}}{\gamma_0}\Big) \right] \right> 
        \\
        &= i \left< \Delta\eta_z e^{i\Delta \psi} \left[ \Omega_{e0}\Big( \frac{1}{\gamma} -\frac{1}{\gamma_0} \Big) + k \Big(\frac{\eta_z}{\gamma} -\frac{\eta_{z0}}{\gamma_0}  \Big) \right] \right> 
        \\
        &\simeq  i \bigg< \Delta \eta_z e^{i\Delta\psi} \Big[ \Omega_{e0}\Big( \gamma_0^{-1} + \Gamma_{-1} \Delta\eta_z +\tfrac{1}{2} \Gamma_{-2}\Delta\eta_z^2 - \gamma_0^{-1} \Big) 
        \\
        & \qquad \qquad + k \eta_z \Big( \gamma_0^{-1} + \Gamma_{-1} \Delta\eta_z +\tfrac{1}{2} \Gamma_{-2}\Delta\eta_z^2 \Big) - k \eta_{z0} \gamma_0^{-1}\Big] \bigg>
        \\
        &\simeq i \bigg <  e^{i\Delta\psi} \Big[ \Delta\eta_z^2\Big( \Gamma_{-1} \Omega_{e0} + k\gamma_0^{-1} + k \Gamma_{-1} \eta_{z0} \Big) 
        \\
        & \qquad \qquad + \Delta\eta_z^3 \Big( \tfrac{1}{2} \Omega_{e0}\Gamma_{-2} + k \Gamma_{-1} + \tfrac{1}{2} k \eta_{z0} \Gamma_{-2} \Big)\Big] \bigg> 
        \\
        &\simeq - i h_c \left< \Delta \eta_z^2 e^{i\Delta \psi} \right>. 
    \end{aligned}
\end{equation}
Combining both terms, 
\begin{equation}
    \dot Y \simeq i u A - i h_c \left< \Delta \eta_z^2 e^{i\Delta \psi} \right>
\end{equation}
Thus, to order $ \langle \Delta \eta_z^2e^{i \Delta\psi} \rangle $ the collective variable equations are,
\begin{align}
    \dot{A} &\ = i g X - i \delta A
    \\
    \dot{X} &\ \simeq - i h_c Y + i \zeta_0 \left< \Delta \eta_z^2 e^{i \Delta \psi} \right> 
    \\
    \dot{Y} &\ \simeq i u A -i h_c \langle \Delta \eta_z^2 e^{i \Delta\psi}\rangle
\end{align}
Since $\zeta_0/h_c \sim 10^{-10}$ for typical magnetospheric parameters such as those considered here, the primarily nonlinear contribution to the equations comes from the second term in $\dot Y$. Thus, 
\begin{align}
    \dot{A} &\ = i g X - i \delta A \label{Eqn: Coll Var Unsimp 1}
    \\
    \dot{X} &\ \simeq - i h_c Y \label{Eqn: Coll Var Unsimp 2}
    \\
    \dot{Y} &\ \simeq i u A -i h_c \langle \Delta \eta_z^2 e^{i \Delta\psi}\rangle \label{Eqn: Coll Var Unsimp 3}
\end{align}
To obtain a closed set of collective variable equations one must eventually apply a closure condition, else continually define higher orders of $\langle \Delta \eta_z^m e^{i \Delta\psi}\rangle$. If we assume moments higher than $m=1$ vanish, then we obtain the \textit{linear} collective variable equations of the free-electron laser model as they appear in \citet{soto-chavez_chorus_2012}, 
\begin{align}
    \dot{A} &\ = i g X - i \delta A \label{Eqn: Linear CVE 1}
    \\
    \dot{X} &\ \simeq - i h_c Y \label{Eqn: Linear CVE 2}
    \\
    \dot{Y} &\ \simeq i u A. \label{Eqn: Linear CVE 3}
\end{align}
To determine the corresponding nonlinear equations, following \citet{bonifacio_collective_1986}, we obtain closure by assuming all moments higher than $m=2$  vanish, and employing the factorization assumption, 
\begin{equation}
    \langle (\Delta \eta_z - \langle \Delta \eta_z \rangle)^2 e^{i \Delta\psi} \rangle \simeq \langle (\Delta \eta_z - \langle \Delta \eta_z \rangle)^2 \rangle \langle e^{i \Delta\psi} \rangle,
\end{equation}
so that we may write, 
\begin{equation} \label{Eqn: Z Unsimplified}
        \langle \Delta \eta_z^2 e^{i\Delta \psi} \rangle \simeq X\left(\langle\Delta\eta_z^2\rangle-2\langle\Delta\eta_z\rangle^2\right)+2Y\langle\Delta\eta_z\rangle. 
\end{equation}
The system admits the following conservation relations which can be used to determine $\langle \Delta \eta_z \rangle$ and $\langle \Delta \eta_z^2 \rangle$, 
\begin{align}
    P_0 &\ = \langle\Delta\eta_z\rangle+ \frac{u}{g} \left|A\right|^2 \label{Eqn: Cons 1}
    \\
    H_0 &\ = \frac{\langle \Delta \eta_z^2 \rangle}{2} - \frac{u}{h_c} \left(A^*X+X^*A\right) + \frac{u \delta}{g h_c} \left| A \right|^2 \label{Eqn: Cons 2}
\end{align}
where $P_0$ and $H_0$ are constants. Their constancy can be shown directly by taking their time derivatives, e.g.
\begin{equation}
    \begin{aligned}
        \dot P_0 &= \langle \dot{\Delta \eta_z}\rangle+ \frac{u}{g} \frac{d}{dt}  \Big(A^* A \Big) 
        \\
        &= \langle \dot{\eta_z}\rangle+ \frac{u}{g} \Big(\dot A^* A + \dot A A^* \Big)
        \\
        &=  \left< i u b e^{-i \psi} - i u b^* e^{i \psi} \right> + \frac{u}{g} \Big[ (- i g X^* + i \delta A^*)A + (i g X - i \delta A)A^* \Big]
        \\
        &= \left< i u b e^{-i \psi_0} e^{i \psi_0} e^{-i \psi} - i u b^* e^{i \psi_0} e^{-i \psi_0} e^{i \psi} \right> + \frac{u}{g} \Big[ - i g X^*A+igXA^* \Big]
        \\
        &= \left< i u A e^{-i \Delta \psi} - i u A^* e^{i \Delta \psi} \right>  + \Big[ - i u X^*A+iuXA^* \Big]
        \\
        &= i u A X^* - i u A^*X - i u X^*A + i u X A^*
        \\
        &= 0
    \end{aligned}
\end{equation}
where we have used \cref{Eqn: EOM 2},  \cref{Eqn: AXY Def 1}, and \cref{Eqn: Coll Var Unsimp 1}. Hence, $P_0$ is an exact constant of equations \eqref{Eqn: EOM 1} - \eqref{Eqn: EOM 3} and reflects the conservation of momentum of the particles and field. Conversely, $H_0$, which is associated with the existence of a Hamiltonian for the system, is an exact constant of the nonlinear collective variable equations and can be derived from \eqref{Eqn: Coll Var Unsimp 1} - \eqref{Eqn: Coll Var Unsimp 2}. Using these two constants, one obtains, 
\begin{align}
    \langle \Delta \eta_z^2 e^{i\Delta \psi} &\rangle \simeq (2H_0-2\frac{u^2}{g^2}|A_0|^4)X + 2\frac{u}{g}|A_0|^2Y \\ 
    &\ + (4 \frac{u^2}{g^2}|A_0|^2-2\frac{u \delta}{g h_c})|A|^2X - 2\frac{u}{g}|A|^2Y + 2\frac{u}{h_c}(A^*X+X^*A)X-2\frac{u^2}{g^2}|A|^4X . \nonumber
\end{align}
The two linear terms constitute a small correction to the exponential growth regime and can be dropped. Similarly, one may omit the first term in the coefficient of $|A|^2X$.  Finally, substituting the expression into equations \eqref{Eqn: Coll Var Unsimp 1} - \eqref{Eqn: Coll Var Unsimp 3}  yields the nonlinear collective variable equations for the free-electron laser model of chorus amplification, 
\begin{gather}
    \dot{A} = i g X - i \delta A \label{Eqn: NLCVE 1}
    \\
    \dot{X} \simeq - i h_c Y \label{Eqn: NLCVE 2}
    \\
    \dot{Y} \simeq i u A  + 2 i \delta\frac{u}{g} |A|^2 X + 2 i h_c\frac{u}{g}|A|^2 Y -2iu(A^*X+X^*A)X + 2ih_c\frac{u^2}{g^2}|A|^4 X. \label{Eqn: NLCVE 3}
\end{gather}
Given the constants, $g,\delta,u, \text{and } h_c$, this simple set of equations allows one to calculate the nonlinear time evolution of the whistler amplitude and phase, and the mean electron momentum and phase.

%% Ch 2 Sec 2.2 %%

\subsection{Derivation of Ginzburg-Landau Equation}

Thus far, we have considered the dynamical behavior of the system in the presence of a single-frequency mode, ignoring its spatial structure. In the presence of spatially varying modes with multiple frequencies it is difficult to obtain a closed set of collective variable equations such as the ones above. However, it was first shown by \citet{cai_ginzburg-landau_1991} by an analogy between the electron beam and an optical fiber that the wave amplitude in the collective variable equations for a FEL can be modeled by a GLE. The analogous equations for chorus amplification (\crefrange{Eqn: NLCVE 1}{Eqn: NLCVE 3}) are of the same form, therefore can also be modeled by a GLE. 

The derivation of the GLE in the FEL context was later given most clearly by \citet{ng_ginzburg-landau_1998}. Giving more details, we follow a similar method, which is based on perturbatively solving \crefrange{Eqn: EOM 1}{Eqn: EOM 3}. To linear order, one has the set \crefrange{Eqn: Linear CVE 1}{Eqn: Linear CVE 3}. If we assume an exponential solution for $A$ of the form
\begin{equation}
    A = A_0 e^{i \lambda_0t}
\end{equation}
where $\lambda_0$ is hereafter denoted the linear growth rate, then we obtain the solutions to the linear collective variable equations
\begin{align}
    A &= A_0 e^{i \lambda_0 t}
    \\
    X &= \frac{\lambda_0 + \delta }{g} A = -\frac{u h_c}{\lambda_0^2} A
    \\
    Y &= -\frac{(\lambda_0+\delta)\lambda_0}{g h_c}A = \frac{u}{\lambda_0}A. \label{Eqn: Linear Soln for Y}
\end{align}
By equating the two expressions for $X$ or $Y$ one obtains the characteristic cubic equation for the linear growth rate
\begin{equation} \label{Eqn: Cubic Equation}
    \lambda_0^3+\delta\lambda_0^2+ u g h_c = 0.
\end{equation}
We continue the perturbative process by substituting the linear solutions into the nonlinear equation for $Y$ and keeping only the first-order nonlinearity (i.e. dropping the $X|A|^4$ term). One obtains
\begin{equation}
    \begin{aligned}
        \dot Y &\simeq i u A + 2 i \delta \frac{u}{g}|A|^2X + 2 i h_c \frac{u}{g} |A|^2 Y - 2 i u(A^*X+X^*A)X
        \\
        &\simeq i u A + 2 i \delta \frac{u}{g} |A|^2 \Big( -\frac{u h_c}{\lambda_0^2} A \Big)+2 i h_c\frac{u}{g}|A|^2\Big( \frac{u}{\lambda_0} A \Big) 
        \\
        &\phantom{\simeq i u A + } - 2iu\bigg[ A^* \Big( -\frac{u h_c}{\lambda_0^2} A \Big) + \Big( -\frac{u h_c}{(\lambda_0^2)^*} A^* \Big)A \bigg]\Big( -\frac{u h_c}{\lambda_0^2} A \Big)
        \\
        & \simeq i u A + 2iu \bigg( -\frac{u h_c \delta}{g \lambda_0^2} + \frac{u h_c}{g \lambda_0} - \frac{u h_c^2}{\lambda_0^4} - \frac{u h_c^2}{\lambda_0^2(\lambda_0^2)^*)} \bigg)|A|^2A
    \end{aligned}
\end{equation}
From \cref{Eqn: Linear Soln for Y} we also know $\dot Y \simeq (u/\lambda_0)\dot A$. Substituting this on the left-hand side yields,
\begin{equation}
        \frac{u}{\lambda_0}\dot A \simeq i u A + 2iu \bigg( -\frac{u h_c \delta}{g \lambda_0^2} + \frac{u h_c}{g \lambda_0} - \frac{u h_c^2}{\lambda_0^4} - \frac{u h_c^2}{\lambda_0^2(\lambda_0^2)^*)} \bigg)|A|^2A.
\end{equation}
After simplifying, one obtains
\begin{equation} \label{Eqn: SLE}
        \dot A \simeq i \lambda_0 A + i\beta|A|^2A
\end{equation}
where we have defined the coefficient\footnote{This coefficient, resulting from the lengthy algebra outlined above, has been verified by computer algebra.}
\begin{equation} \label{Eqn: Beta}
    \beta \equiv -2 u h_c \lambda_0 \left(  -\frac{1}{g\lambda_0} + \frac{ \delta}{g \lambda_0^2} 
    + \frac{u h_c}{\lambda_0^4} + \frac{u h_c}{|\lambda_0|^4}\right).
\end{equation}
\Cref{Eqn: SLE} is the SLE mentioned above. It provides a simplified model for the amplitude behavior, describing initial exponential growth followed by saturation at a value of $\langle|A_{\text{sat}}|\rangle= \sqrt{-\lambda_i/\beta_i}$,  where $\lambda_r +i \lambda_i \equiv \lambda_0$ and $\beta_r + i \beta_i \equiv \beta$. It is important to note that $\lambda_0$, being derived from the solution to the cubic equation, has three solutions. The stability analysis of \cref{Eqn: Cubic Equation} reveals one real root and a pair of complex conjugate roots \cite{soto-chavez_chorus_2012}. Given that the form of the exponential solutions is e.g. $A = A_0 e^{i\lambda_0 t}$, the complex root with the negative imaginary part will dominate the growth. 

\begin{figure}[!ht]
    \centering
    \includegraphics[width=0.71\linewidth]{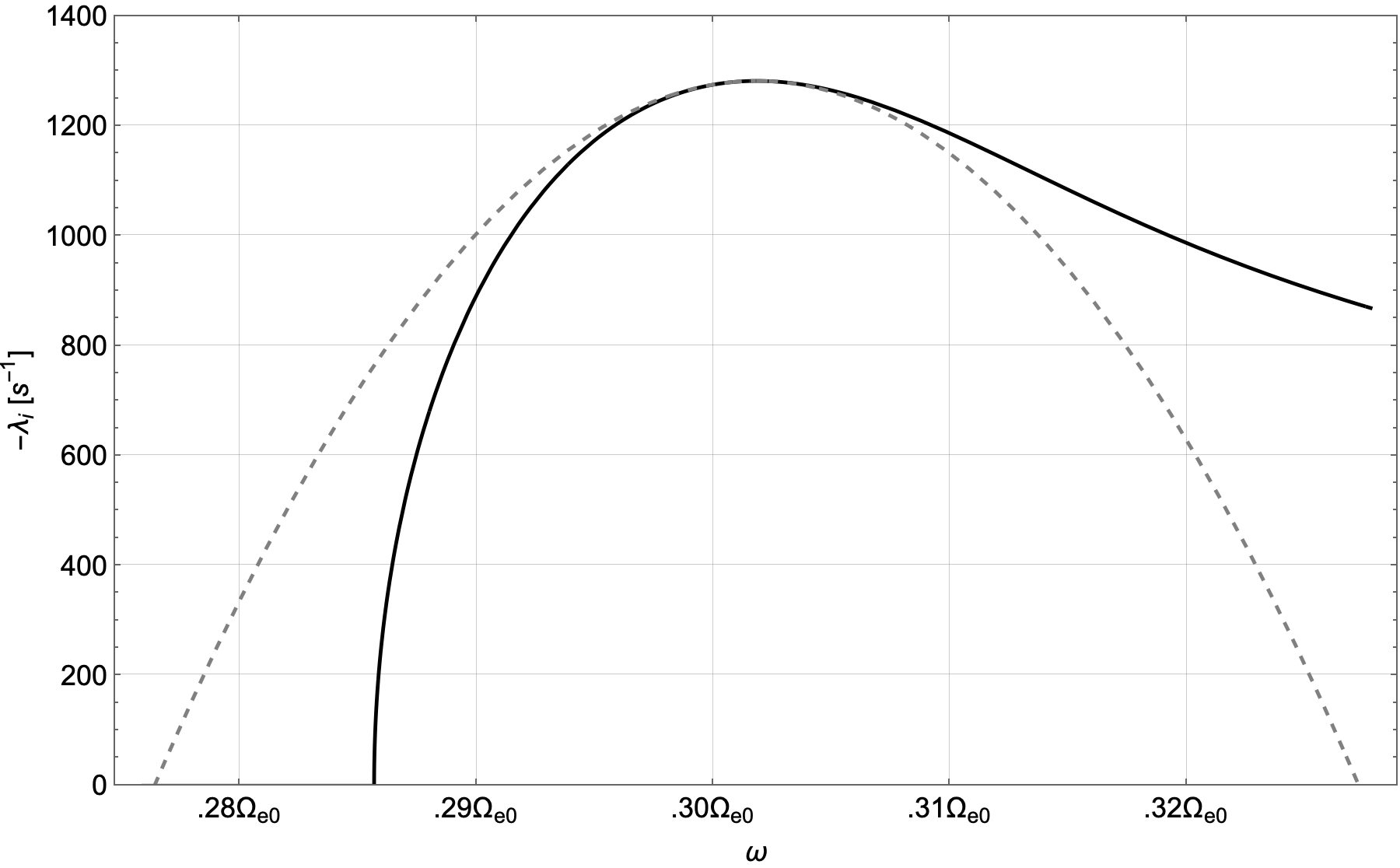}
    \caption{The growth rate (solid line) as a function of the whistler frequency, $-\lambda_i(\omega)$, and its parabolic approximation (dashed line) near the resonance frequency. }
    
    \label{Fig: Ch 2 Lambda Appx}
\end{figure}

Next we derive the GLE by extension of the SLE. First we allow for a spectrum of frequencies near a reference frequency $\omega_s$. Defining $\Delta\omega = \omega-\omega_s$ one has 
\begin{equation} \label{Eqn: lambda taylor expansion}
    \begin{aligned}
        \lambda_0(\omega) &\simeq \lambda_0(\omega_s) + \frac{\partial \lambda_0(\omega_s)}{\partial \omega} \Delta \omega + \frac{1}{2} \frac{\partial^2 \lambda_0(\omega_s)}{\partial \omega^2} \Delta\omega^2
        \\
        & \simeq \lambda_0(\omega_s) + \frac{\partial \lambda_0(\omega_s)}{\partial \omega} i\frac{\partial}{\partial t} + \frac{1}{2} \frac{\partial^2 \lambda_0(\omega_s)}{\partial \omega^2} i^2\frac{\partial^2}{\partial t^2}
        \\
        & \simeq \lambda_0(\omega_s) + i\frac{\partial \lambda_0(\omega_s)}{\partial \omega} \frac{\partial}{\partial t}  -\frac{\alpha}{2} \frac{\partial^2}{\partial t^2}
    \end{aligned}
\end{equation}
where
\begin{equation}
    \alpha \equiv \frac{\partial^2 \lambda_0(\omega_s)}{\partial \omega^2}
\end{equation}
and we have used the relation between the frequency and the time derivative of a Fourier mode. This order of approximation is justified by the parabolic shape of $\lambda_i$, at the maxima near the cyclotron resonance frequency as shown in \Cref{Fig: Ch 2 Lambda Appx} for the typical magnetospheric parameters considered in this work. 

Next, we relax the previous constraint that the whistler amplitude $B_w\propto|A|$ and phase shift $\phi$ are spatially homogeneous, so that 
\begin{equation} \label{Eqn: Derivative Extension}
    \begin{aligned}
        \frac{d}{dt} = \frac{\partial}{\partial t} + \frac{dz}{dt} \frac{\partial}{\partial z} = \frac{\partial}{\partial t} + v_g \frac{\partial}{\partial z}
    \end{aligned}
\end{equation}
where we have used the fact that the speed of the wave packet is $dz/dt = d\omega/dk = v_g$. Using \cref{Eqn: Derivative Extension} and \cref{Eqn: lambda taylor expansion} in equation \cref{Eqn: SLE} one obtains, 
\begin{gather}
    \bigg(\frac{\partial}{\partial t} +v_g \frac{d}{dz}\bigg) A 
    - i \bigg(\lambda_0(\omega_s) + i \frac{\partial \lambda_0(\omega_s)}{\partial \omega} \frac{\partial}{\partial t} - \frac{\alpha}{2} \frac{\partial^2}{\partial t^2}\bigg)A - i \beta |A|^2A = 0.
\end{gather}
After rearranging, one obtains the chorus GLE 
\begin{equation} \label{Eqn: GLE Full}
    v_g \pdv{A}{z} = i \lambda_0(\omega_s)A - \mu \pdv{A}{t} - i \frac{\alpha}{2} \pdv[2]{A}{t}+ i \beta |A|^2 A.
\end{equation}
where we have defined
\begin{equation} \label{Eqn: mu definition}
    \mu \equiv 1 + \frac{\partial \lambda_0(\omega_s)}{\partial\omega}.
\end{equation}
This can be written in a simpler form which depends on just two complex constants. For convenience, we define $\mu_r+i\mu_i\equiv\mu$, $\alpha_r+i\alpha_i\equiv\alpha$, denote division by the group velocity with an overbar, e.g. $\bar\lambda_r\equiv\lambda_r/v_g$, and define $c_1=-\alpha_r/\alpha_i$ and $c_2 = -\beta_r/\beta_i$. Then by applying the transformations $A(z,t) = \Phi_0 \Phi(\zeta,\tau)\exp i[K_0 \zeta + \Omega_0 \tau]$, $\zeta\equiv z/z_0$, $\tau\equiv t/t_0+z/v_0t_0$, where 
\begin{equation}
\begin{aligned}
    z_0 &= 1/(-\bar{\lambda}_i+\bar{\mu}_i^2/2\bar{\alpha}_i),& t_0^2 &= \bar{\alpha}_i z_0/2, \\
    v_0 &= 1/(-\bar{\mu}_r +\bar{\alpha}_r \bar{\mu}_i /\bar{\alpha}_i),& \Phi_0^2 &= 1/z_0\bar{\beta}_i, \\
    K_0 &= z_0(\bar{\lambda}_r - \bar{\alpha}_r\bar{\mu}_i^2/2\bar{\alpha}_i^2),& \Omega_0 &= \bar{\mu}_i t_0 / \bar{\alpha}_i
\end{aligned}
\end{equation}
one obtains the GLE in standard form
\begin{equation} \label{Eqn: GLE Std.}
    \pdv{\Phi}{\zeta} = \Phi + (1+ic_1)\pdv[2]{\Phi}{\tau} -(1+ic_2)\abs{\Phi}^2\Phi.
\end{equation}

%% Ch 2 Sec 3 %%

\section{Discussion}

\begin{figure}[!ht]
    \centering
    \includegraphics[width=0.71\linewidth]{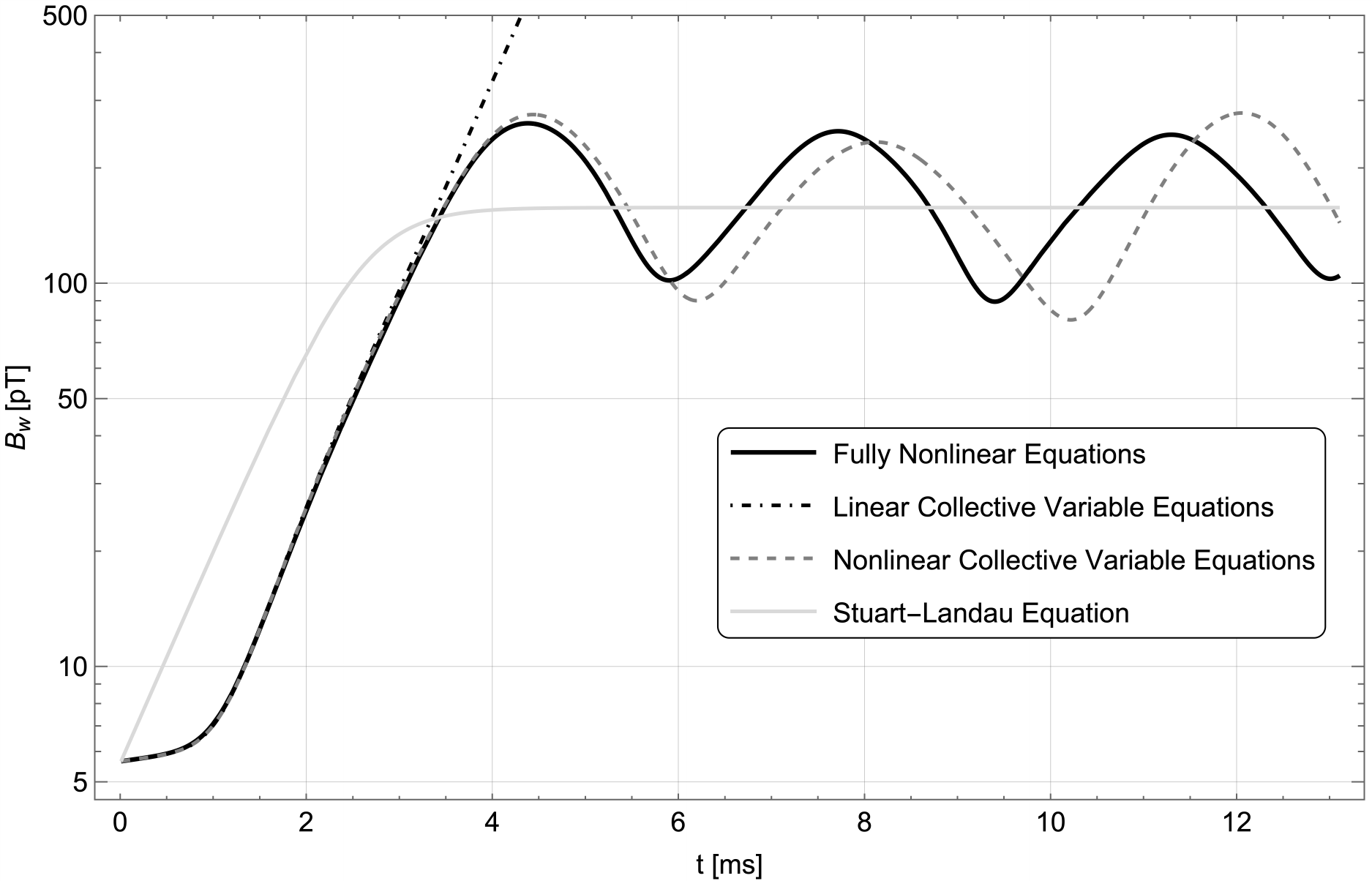}
    \caption{A comparison between the fully nonlinear original 2$N$+2 dynamical equations (Eqns. \eqref{Eqn: EOM 1} - \eqref{Eqn: EOM 3}), the linear collective variable equations (Linearized Eqns. \eqref{Eqn: NLCVE 1} - \eqref{Eqn: NLCVE 3}) appearing in the work of \citet{soto-chavez_chorus_2012}, the nonlinear collective variable equations (Eqns. \eqref{Eqn: NLCVE 1} - \eqref{Eqn: NLCVE 3}), and the Stuart-Landau Equation (\cref{Eqn: SLE}) for typical magnetospheric conditions. The nonlinear collective variable equations agree with the linear collective variable equations in the exponential growth regime, and agree with the fully-nonlinear equations in the post-saturation regime.}
    
    \label{Fig: Figure 1}
\end{figure}

In the single-mode case, the general behavior of the wave amplitude is exponential growth followed by saturation and post-saturation oscillations. This is shown in \cref{Fig: Figure 1} for typical magnetospheric conditions, such as those used in the original FEL model analysis by \citet{soto-chavez_chorus_2012}.  For completeness, we restate the conditions here: a whistler with a frequency of $\omega_s = 0.3 \Omega_{e0}$ (near the resonance frequency occurring at $\delta=0$), wavelength of 8131 km, and initial amplitude of 5.7 pT interacting with a monoenergetic beam of electrons with initial parallel velocities of $v_{z0}=-0.157 c$, constant perpendicular velocities of $v_{\perp0}=0.68c$, initial bunching parameter of $X_0 \approx 0.001$, and plasma frequency of $\omega_{pr} = 10^3 \text{rad/s}$ located at the magnetic equator on the L=4 shell where the background field is 0.5 \textmu T. 

Figure \ref{Fig: Figure 1} shows exponential growth at a rate of $|\lambda_i| \sim$ 10\textsuperscript{3} s\textsuperscript{-1} for a few milliseconds with a saturation amplitude around 250 pT followed by amplitude modulations of width $\sim$2 ms and period $\sim$3.5 ms. Away from the narrow resonance at $\delta=0$, the linear growth rates are of the order of hundreds of s\textsuperscript{-1}. These predictions are in good agreement with in situ satellite measurements, which indicate exponential growth at a rate of up to a few hundred s\textsuperscript{-1} followed by a peak amplitude at typical values of a few hundred pT \cite{santolik_new_2008}.  Furthermore, these measurements indicate that amplitude peaks can be followed by periods of decay and further growth, resulting in amplitude oscillations with a duration of a few milliseconds to a few tens of milliseconds. Such amplitude modulations are common, having been observed in many other cases \cite{dubinin_coherent_2007, mozer_whistlers_2021, santolik_fine_2014, li_modulation_2011}. Last, one of the dominant motivations for the development of this laser-like model is the well-known fact that chorus modes are highly phase coherent \cite{agapitov_chorus_2017}, which is attributed to their resonant interaction with phase bunched (coherent) electrons. This is reflected in our model by the phase bunching parameter $X$ growing from its initially near-random state to a peak value around 0.8 in lockstep with the amplitude growth. 

\begin{figure}
    \centering
    \includegraphics[width=1\linewidth]{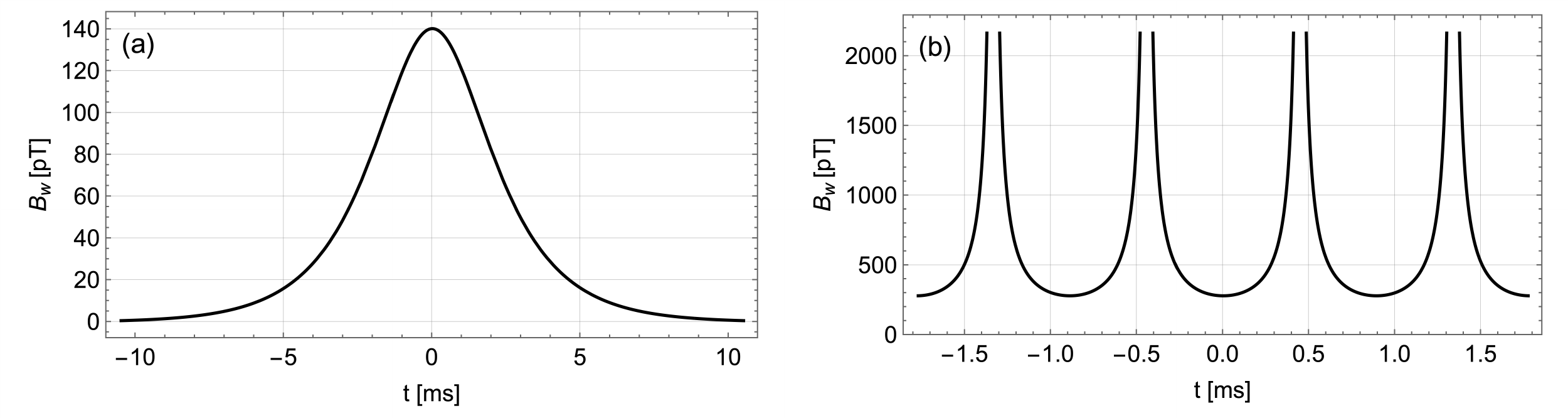}
    \caption{Solitary wave solutions of the Ginzburg-Landau equation (\cref{Eqn: GLE Full}). (a) The solitary pulse corresponding to \Cref{Eqn: Single Solitary Wave}. (b) The periodic train of solitary waves with finite-time singularities corresponding to \Cref{Eqn: Periodic Solitary Wave}.} 
    
    \label{fig: Figure 2}
\end{figure}

The multi-mode spatially-varying chorus behavior is governed approximately by the GLE. The bandwidth of the resulting wave packets can be estimated using the parabolic approximation for the growth rate employed in the derivation of the GLE, resulting in the expression, $\delta\omega\approx2\sqrt{-2\lambda_i/\alpha_i}$, or $\delta\omega \approx$ 0.05 $\Omega_{e0}$ for the parameters considered here. This is consistent with CLUSTER spacecraft observations, which indicate the majority of chorus wave packets have a bandwidth in the range of 0.03 $\Omega_{e0}$ to 0.2 $\Omega_{e0}$, with lower values occurring more often near the magnetic equator \cite{santolik_frequencies_2008}. Furthermore, it is well known that the GLE admits special solitary wave solutions. This predicts that, despite being composed of a spectrum of frequencies - each with distinct growth rates and phase velocities - the amplitude envelope of a wave packet can propagate with a definite shape and velocity, as first described by Russell \cite{russell_report_1844} for the case of shallow water waves. Such solutions have been derived in general \cite{nozaki_exact_1984, cariello_painleve_1989}, and discussed more particularly for the case of free-electron lasers \cite{cai_ginzburg-landau_1991}. In the latter case, in which the GLE is derived from collective variable equations of the same form as appear here, the pulse widths have been compared favorably to data from several free-electron laser experiments. The solitary wave solutions to \cref{Eqn: GLE Std.} are pulses of the form, 
\begin{equation}
    \Phi = \frac{Q e^{-i\Omega\zeta}}{(2 \cosh K\tau )^{1+i\sigma}},
\end{equation}
where, 
\begin{equation}
    |Q| = 2\left(1+\frac{c_1 \sigma-1}{\sigma^2+2c_1\sigma-1}\right)^{1/2},
    \Omega=c_1 - \frac{2\sigma(1+c_1)}{\sigma^2+2c_1\sigma-1},
    K = \left(\frac{1}{\sigma^2+2c_1\sigma-1}\right)^{1/2},
\end{equation}
and $\sigma$ satisfies the quadratic equation, 
\begin{equation}
    \sigma^2-3\frac{c_1c_2+1}{c_1-c_2}\sigma-2=0.
\end{equation}
By transforming back to the original field and coordinates, we obtain the solitary wave solutions to \cref{Eqn: GLE Full} in physical space, 
\begin{equation}
    A = \frac{\Phi_0Q}{\left[2\cosh K(\frac{t}{t_0}+\frac{z}{v_0 t_0})\right]^{1+i\sigma}}\exp{i\left[ z\left( \frac{K_0}{z_0} + \frac{\Omega_0}{v_0 t_0} - \frac{\Omega}{z_0}\right) +t\left( \frac{\Omega_0}{t_0} \right) \right]}.
\end{equation}
For the magnetospheric parameters considered here, the roots of the equation for $\sigma$ are real, but result in one purely real $K$ and one purely imaginary $K$. In the former case, where $K = K_r \in \mathbb{R}$, at the origin one obtains the single pulse, 
\begin{equation} \label{Eqn: Single Solitary Wave}
    |A(t)| = \frac{|\Phi_0 Q|}{2\cosh{(K_r t/t_0)}}.
\end{equation}
In the latter case, where $K = K_i \in i\mathbb{R}$, at the origin one obtains multiple pulses with periodic singularities,
\begin{equation} \label{Eqn: Periodic Solitary Wave}
    |A(t)| = \frac{|\Phi_0Q|}{|2\cos{(|K_i| t/t_0)}|},
\end{equation}
where we have used the approximation $|\sigma|\ll1$. Plots of equations \eqref{Eqn: Single Solitary Wave} and \eqref{Eqn: Periodic Solitary Wave} are shown in figure \ref{fig: Figure 2}. The single pulse is gaussian-like, with a width given by $\Delta t = 2 t_0/K_r$, which is approximately $\Delta t \simeq \sqrt{-2\alpha_i \sigma^2/ \lambda_i}$ in the limit $\sigma \gg 1$.  Substituting the parameters used here yields a peak amplitude of 140 pT and a width of $\Delta t \approx$ 3.5 ms. In contrast, the singular periodic solution consists of a train of narrow spikes, with widths given by $\Delta t = 2 t_0/|K_i| \approx$ 0.5 ms, which is approximately  $\Delta t \simeq \sqrt{-2\alpha_i/ \lambda_i}$ in the limit $\sigma \ll 1$. The separation between neighboring singularities is given by $\Delta T = \pi t_0/|K| \approx$ 0.9 ms, which is approximately $\Delta T \simeq \pi \sqrt{-\alpha_i / 2\lambda_i}$ for $\sigma \ll 1$. Although the validity of the periodic solution (with its finite-time singularities) is limited and the solutions will be regularized by effects outside the realm of validity of our model, it nevertheless predicts a qualitatively similar sequence of millisecond timescale amplitude spikes and a tendency for spike formation. In fact, both solutions exhibit modulation timescales comparable to those of the single‐mode case, in agreement with the observational evidence discussed above. Furthermore, in the context of magnetospheric lion roars, \citet{dubinin_coherent_2007} reported that bursts of whistler emissions - also excited by a local population of resonant electrons - exhibit amplitude variations consistent with solitary wave behavior. The associated wave turbulence was found to consist of many nearly monochromatic, circularly polarized wave packets, with field strength variations resembling solitary structures. Given that chorus arises from a similar mechanism, it is plausible that chorus modes also exhibit such solitary wave behavior. 

Finally, we note some limitations of the present model and possible directions for future research. First, the single-mode treatment - which results in the reduced nonlinear equations (Eqns. \eqref{Eqn: NLCVE 1} - \eqref{Eqn: NLCVE 3}) - assumes a whistler wave with a fixed frequency and a time-dependent phase. Despite the tendency of chorus modes to chirp in frequency even in the absence of background inhomogeneities \cite{omura2011triggering}, here we do not investigate the physics of the frequency variations contained within the time-dependent phase. In general, a multi-mode approach, similar to that employed in our derivation of the GLE, may also be capable of modeling chirping behavior. However, since the SLE averages out the post-saturation oscillations indicative of resonant electron trapping - one of the mechanisms associated with frequency chirping - the degree to which the GLE, as presented here, can model chorus chirping is uncertain. The investigation of chirping in the context of the FEL model is a promising direction for future research. Since the FEL analogy works in both directions, such an investigation could also provide insight into frequency chirping in free-electron lasers. Next, while equations \eqref{Eqn: NLCVE 1} - \eqref{Eqn: NLCVE 3} offer a straightforward route to the analysis of both wave and particle behavior, this work focuses primarily on the wave behavior, leaving the investigation of the model's implications for particle dynamics to future work. In addition, we note that our derivation of the GLE, being based on an extension of single-mode spatially-independent equations could perhaps be derived more directly as a limiting case of a more general treatment. Last, we note the interesting possibility for the exploration of the FEL model in related circumstances, such as ion-cyclotron waves in the magnetosphere, or toroidal configurations, where the excitation of whistler waves by runaway electrons has recently been observed \cite{spong_first_2018}. 

%% Ch 2 Sec 4

\section{Conclusion}

In this work, we investigated nonlinear aspects of whistler-mode chorus amplification in the magnetosphere using the free-electron laser model. In the single-mode case, we derived nonlinear collective variable equations for the system, predicting exponential growth followed by saturation and post-saturation amplitude oscillations consistent with observational data. Next, we considered a packet of whistlers with a spectrum of frequencies and spatially dependent amplitudes, and found that the wave behavior is approximately governed by the complex Ginzburg-Landau equation which admits solitary wave solutions. We found that both the single-mode and multi-mode equations predict amplitude modulations on millisecond timescales, consistent with observations. Further exploration into the free-electron laser model, including the newly proposed Ginzburg-Landau equation, its stability, and its observational implications, is the subject of the following chapter. 

%%%%%%%%%%%%%%%
%% Chapter 3 %%
%%%%%%%%%%%%%%%

\chapter{Single-Mode Instabilities} \label{Chapter 3: Single-Mode Instabilities}

This chapter is based on results currently under review at \textit{The Journal of Geophysical Research: Space Physics}, co-authored with Amitava Bhattacharjee \cite{bonham2025ginzburg}. 

\section{Introduction}

In the previous chapter we discussed recent theoretical developments in the free-electron laser (FEL) model of whistler-mode chorus in the magnetosphere. Perhaps of greatest theoretical interest was the finding that when the single frequency theory was extended to accommodate a spectrum of frequencies with spatially varying amplitudes, the whistler amplitude and phase were found to be governed by one of the most well-studied nonlinear equations in physics - the Ginzburg-Landau equation (GLE). Therefore, one can model the nonlinear dynamics of a chorus wave packet using a single partial differential equation, rather than a system of 2$N$+2 coupled nonlinear partial differential equations. While the GLE can describe the rich nonlinear physics that arises from multi-mode interactions, it also permits the existence of, and condensation to, single-mode solutions. In this chapter, we apply insights from the study of single-mode operation in FELs to investigate the linear stability of single modes in the context of magnetospheric chorus. 

Mode condensation is characterized by the collapsing of power from an initially broad spectrum of modes down to a narrow band or single-frequency mode. The potential for mode condensation depends on the relationship between frequency and stability for a given system. In general, the instability conditions for single-mode solutions of the GLE are well-known \cite{aranson_world_2002, garcia-morales_complex_2012}. The calculation has also been repeated in the context of a FEL \cite{ng_ginzburg-landau_1998}, and since the form of the chorus GLE is the same as the FEL GLE, the results can be readily translated. One key insight from the FEL literature is that if the principal mode (the mode with the highest linear growth rate) is unstable, then all modes are unstable, corresponding to the Benjamin-Feir instability \cite{benjamin1967disintegration}. Or conversely, if the principal mode is stable, then there is a continuous band of stable modes surrounding it, with modes outside that band being unstable, corresponding to the Eckhaus instability \cite{eckhaus_studies_1965}. 

The work is organized as follows. First, we introduce the GLE and its single-mode solutions. We then outline the process for obtaining the inequalities that express the linear stability conditions for these solutions. Next, we introduce a perturbation approximation for the linear growth rate of the whistler, which is one of the key physical parameters which determines the stability behavior. From this perturbation approximation, we show that the mode with the highest linear growth rate is always stable, hence the model predicts no Benjamin-Feir instability. Again employing the perturbation approximation for the linear growth rate, we obtain a simple expression for the Eckhaus instability condition which immediately allows one to compute the width of the Eckhaus stable frequency band. Importantly, we find that both instability conditions are approximately independent of the specific parameters of the system insofar as the basic assumptions of the FEL model are met, including that the resonance frequency must be near the maximum linear growth rate frequency. Last, we perform numerical simulations to support these analyses, and demonstrate the existence of mode condensation. 

In short, this study provides a new perspective on the development of narrow band chorus from a noisy spectrum through the process of mode condensation. This can be viewed as the tendency for modes outside of the Eckhaus stability band to transition to modes within the stability band. In addition, this work shows that, under suitable assumptions, the stability conditions and therefore bandwidth estimates can be simplified to a trivial form. Ultimately, these results provide yet another interesting example of the insights which can be gained into the problem of magnetospheric chorus by drawing upon the FEL model and existing FEL literature. 

\section{Ginzburg-Landau Equation}

So that this chapter may be relatively self-contained, we briefly review the GLE for the evolution of chorus due to resonant interactions with radiation belt electrons, 
\begin{equation} \label{Eqn: GLE}
    \frac{\partial \Phi}{\partial \zeta} = \Phi + (1+ic_1)\frac{\partial^2 \Phi}{\partial \tau^2} -(1+ic_2)|\Phi|^2\Phi,
\end{equation}
where $c_1 = -\alpha_r/\alpha_i$ and $c_2 = -\beta_r/\beta_i$. This form of the GLE, dependent on just two real constants, is the most simple and standard, making it well suited for analytical manipulation and comparison with existing literature \cite{aranson_world_2002}. The parameters $\alpha_r +i\alpha_i \equiv \alpha$ and $\beta_r + i \beta_i \equiv \beta$ are defined by, 
\begin{align}
    \alpha &\equiv \left. \frac{\partial^2 \lambda_0(\omega)}{\partial\omega^2} \right|_{\omega_s} \\
    \beta &\equiv \left. -2 u h_c \lambda_0 \left(  -\frac{1}{g\lambda_0} + \frac{ \delta}{g \lambda_0^2} + \frac{u h_c}{\lambda_0^4} + \frac{u h_c}{|\lambda_0|^4}\right) \right |_{\omega_s}. \label{Eqn: beta}
\end{align}
Physically, $\alpha$ is the curvature (second derivative) of the linear growth rate of the whistler, $\lambda_0(\omega) =\lambda_0$. The linear growth rate itself is the dominant root (that with the largest negative imaginary part) of the characteristic equation of the linearized collective variable equations (\crefrange{Eqn: NLCVE 1}{Eqn: NLCVE 3}),
\begin{equation}
    \lambda_0^3+\delta\lambda_0^2+ug h_c = 0.
\end{equation}
The constants $\delta$, $u$, $g$, and $h_c$ (derived in \cref{Sec: Derivation of 2N+2 Equations}) are defined as, 
\begin{equation}
\begin{aligned}
    \delta &\equiv k (v_{z0}-v_r) \\
    u      &\equiv v_{\perp 0}/2 \\
    g      &\equiv \frac{s \Omega_{e0} \omega_{pr}^2 u}{k c^2} \\
    h_c    &\equiv -\frac{k}{\gamma_0}(1- \frac{\gamma_{\perp 0}^2\eta_{z0}^2}{\gamma_0^2 c^2} -\frac{\gamma_{\perp 0}^2\Omega_{e0}\eta_{z0}}{\gamma_0^2 c^2 k}),
\end{aligned}
\end{equation}
where $v_r \equiv (\omega-\Omega_{e0}/\gamma_0)/k$ is the gyroresonance velocity, $s\equiv \omega/(\Omega_{e0}-\omega)$, $\Omega_{e0} \equiv eB_0/m_e$ is the electron cyclotron frequency due to the background geomagnetic field,  $\omega_{pr} \equiv (n_r e^2/\epsilon_0 m_e)^{1/2}$ is the resonant electron plasma frequency, $\gamma_0 \equiv (1 - \bm v_0^2/c^2)^{-1/2}$ is the resonant electron initial Lorentz factor, $\gamma_{\perp 0} \equiv (1 - v_{\perp 0}^2/c^2)^{-1/2}$ is the resonant electron perpendicular Lorentz factor, $\eta_{z 0} \equiv \gamma_0 v_{z0}$ is the resonant electron initial proper velocity parallel to $\bm B_0$. Lastly, $n_r$ is the number density of resonant electrons, $e$ is the elementary charge, $\epsilon_0$ is the permittivity of free space, $m_e$ is the mass of the electron, and $c$ is the speed of light in vacuum. Importantly, $\delta$, $g$, and $h_c$ all depend implicitly on $\omega$ through their dependence on $k$. 

In exchange for the simplicity of equation \eqref{Eqn: GLE}, one must apply two transformations to recover the physical field, $\Phi(\zeta,\tau) \rightarrow A(z,t) \rightarrow \bm{B}_w(z,t)$, first converting from the rescaled collective variable in scaled space and time coordinates, $\Phi(\zeta,\tau)$, then to the standard collective variable in physical space and time coordinates, $A(z,t)$, and finally to the whistler magnetic field, $\bm B_w$. To convert from the rescaled collective variable in scaled space and time coordinates, $\Phi(\zeta,\tau)$, to the standard collective variable in ordinary coordinates, $A(z,t)$, we apply the transformations, 
\begin{align}
    A(z,t) &= \Phi_0 \Phi(\zeta,\tau)\exp i[K_0 \zeta + \Omega_0 \tau] \\
    \zeta &= z/z_0 \\
    \tau &= t/t_0+z/v_0t_0.
\end{align}
The scaling constants are given by, 
\begin{equation}
\begin{aligned}
    z_0 &= 1/(-\bar{\lambda}_i+\bar{\mu}_i^2/2\bar{\alpha}_i),& t_0^2 &= \bar{\alpha}_i z_0/2, \\
    v_0 &= 1/(-\bar{\mu}_r +\bar{\alpha}_r \bar{\mu}_i /\bar{\alpha}_i),& \Phi_0^2 &= 1/z_0\bar{\beta}_i, \\
    K_0 &= z_0(\bar{\lambda}_r - \bar{\alpha}_r\bar{\mu}_i^2/2\bar{\alpha}_i^2),& \Omega_0 &= \bar{\mu}_i t_0 / \bar{\alpha}_i,
\end{aligned}
\end{equation}
where $ \mu \equiv 1 + \left .\frac{\partial \lambda_0}{\partial\omega} \right|_{\omega_s}$, the subscripts denote real and imaginary parts, the overbars denote division by the group velocity, e.g. $ \bar \alpha_i \equiv \alpha_i /v_g$, and the group velocity can be calculated from the cold whistler dispersion relation. Finally, the whistler field may be recovered from $A$ via the relation, $A = (e/m_e)B_w e^{i(\phi-\delta t)}$. Or if we denote the complex phase of $A$ as $\varphi_A$, then the whistler amplitude and phase is fully specified by,
\begin{equation}
\begin{aligned}
    B_w &= \frac{m_e}{e} |A| \\
    \phi &= \varphi_A + \delta t.
\end{aligned}
\end{equation}
\section{Single-Mode Solutions}

It can be verified by direct substitution that the single-frequency mode, 
\begin{equation} \label{Eqn: SM Solution}
    \Phi = \sqrt{1-\omega_0^2} \exp i\left(k_0 \zeta-\omega_0 \tau + \phi_0\right)
\end{equation}
is a solution to equation \eqref{Eqn: GLE}, for any $\omega_0 \in (-1,1)$ where $k_0 = -c_2 - (c_1-c_2)\omega_0^2$. Note, the constant $\phi_0 \in \mathbb R$, which determines the phase at $\zeta = \tau = 0$ is hereafter assumed zero for simplicity. To extract the whistler frequency corresponding to this solution, we transform from $\Phi(\zeta,\tau)$ to $A(z,t)$, then employ the operator identity $i \partial/\partial t = \Delta\omega \equiv (\omega -\omega_s)$ which originates from the Fourier analysis which led to the GLE. Applying this operator to $A$ yields, 
\begin{equation} \label{Eqn: SM Frequency Long}
    \omega = \omega_s - \frac{\mu_i}{\alpha_i} + \frac{\omega_0}{t_0}.
\end{equation}
By Taylor expansion of $\lambda_i(\omega)$ it can be shown that the frequency with the maximum linear growth rate, $\omega_m$, may be written as $\omega_m = \omega_s - \mu_i/\alpha_i + O\left(|\omega_m - \omega_s|^2\right)$ \cite{ng_ginzburg-landau_1998}. This can be understood by noting that the maxima of the parabolic approximation of $\lambda_i(\omega)$ is itself given by, $\omega_m = \omega_s - \mu_i/\alpha_i$, since to second order
\begin{align}
    \lambda_i(\omega) &\simeq \lambda_i(\omega_s) + \lambda_i'(\omega_s)(\omega-\omega_s) + \frac{1}{2}\lambda_i''(\omega_s)(\omega - \omega_s)^2
    \\
    &\simeq \lambda_i(\omega_s) + \mu_i(\omega-\omega_s)+\frac{1}{2}\alpha_i(\omega-\omega_s)^2
\end{align}
and since the maxima is found at $\lambda_i'(\omega_m)=0$ one obtains
\begin{equation}
    \omega_m \simeq \omega_s - \frac{\mu_i}{\alpha_i}.
\end{equation}
The error in the parabolic approximation for lambda is of order $|\omega_m-\omega_s|^3$ so the error in the condition for the maxima, $\lambda_i'(\omega_m)=0$, is of order $|\omega_m-\omega_s|^2$, thus
\begin{equation}
    \omega_m = \omega_s - \mu_i/\alpha_i + O\left(|\omega_m - \omega_s|^2\right).
\end{equation}
Since the GLE derivation requires that the reference frequency, $\omega_s$, is near the maximum linear growth rate frequency, $\omega_m$, the error is small by construction. For example, for typical magnetospheric parameters \cite{omura_theory_2008,lampe2010nonlinear,santolik_spatio-temporal_2003} and $\omega_s=\omega_\delta$ (where $\omega_\delta$ is defined in section \ref{Sec: Stability Bandwidth}) the relative error is of order one part in a thousand. Therefore, the frequency of the single-mode solution is approximately a shift of $\omega_0/t_0$ away from the maximum growth mode and equation \eqref{Eqn: SM Frequency Long} may be written,  
\begin{equation}
    \omega \simeq \omega_m + \frac{\omega_0}{t_0}.
\end{equation}
\subsection{Instability Criteria}

The linear stability of equation \eqref{Eqn: SM Solution} can be analyzed by adding a small perturbation, substituting the perturbed expression into the GLE, then re-solving the GLE to linear order. If the perturbed solution grows exponentially, then the original solution is linearly unstable. This procedure, sketched here, is well known in general, and has also been conducted in particular in the FEL context. When written in standard form, the FEL GLE is identical to the chorus GLE, apart from the numerical values of $c_1$ and $c_2$. Thus, the stability results can be carried over straightforwardly. Here we sketch the analysis, presented for FELs by \citeauthor{ng_ginzburg-landau_1998}, and apply it to the problem of magnetospheric chorus. 

It will be shown from the instability conditions (equations \eqref{Eqn: BF Criterion} and \eqref{Eqn: Eckhaus Criterion}) that if the mode with the maximum linear growth rate (the $\omega_0 = 0$ mode) is unstable then all single modes are unstable, corresponding to the Benjamin-Feir instability. We first sketch this case. To analyze the linear stability of the mode, $\Phi = \exp (-ic_2\zeta)$, we add small perturbations $a(\zeta,\tau) = \tilde{a}(\zeta) \cos(\omega_d \tau)$ and $\theta(\zeta,\tau) = \tilde{\theta}(\zeta) \cos(\omega_d \tau)$ to the amplitude and phase respectively to obtain the perturbed mode $\tilde \Phi$, 
\begin{equation}
    \Phi \rightarrow \tilde{\Phi} = \bigl(1+\tilde{a}(\zeta) \cos(\omega_d \tau) \bigr )\exp{i\left[-c_2 \zeta +\tilde{\theta}(\zeta) \cos(\omega_d \tau)\right]}.
\end{equation}
Using this ansatz in the GLE and linearizing with respect to  $\tilde a(\zeta)$ and $\tilde \theta(\zeta)$ yields the exponential solutions $\tilde a(\zeta) = \tilde a_0\exp(\Lambda \zeta)$ and $\tilde \theta(\zeta) = \tilde \theta_0 \exp (\Lambda \zeta)$, where $\tilde a_0 \equiv \tilde a(0)$,  $\tilde \theta_0 \equiv \tilde \theta(0) $, and the growth rate of the perturbations is, 
\begin{equation}
    \Lambda = -(1+\omega_d^2) \pm \sqrt{1-2 c_1 c_2 \omega_d- c_1^2 \omega_d^4}.
\end{equation}
If $\Lambda$ has a positive real part then the perturbations will grow exponentially and $\Phi$ will be unstable. This yields the Benjamin-Feir-Newell \textit{instability} criterion, 
\begin{equation} \label{Eqn: BF Criterion}
    1 + c_1 c_2 < 0. 
\end{equation}
If the $\omega_0=0$ mode is stable, then there is a band of stable values of $\omega_0$ outside of which all single modes are unstable, known as the Eckhaus instability. To show this, we introduce the aforementioned perturbations to the $\omega_0 \not = 0$ solution, 
\begin{equation} \label{Eqn: Perturbed SM}
    \Phi \rightarrow \tilde{\Phi} = 
    \sqrt{1-\omega_0^2}\Big(1+\tilde{a}(\zeta) \cos(\omega_d \tau) \Big )
    \exp{i\left[k_0 \zeta - \omega_0 \tau +\tilde{\theta}(\zeta) \cos(\omega_d \tau)\right]}.
\end{equation}
As before, linearizing the GLE using this ansatz yields exponential solutions for $\tilde a(\zeta)$ and $\tilde \theta(\zeta)$. The growth rate of low frequency perturbations ($\omega_d \ll 1)$ is given by, 
\begin{equation}
    \Lambda \simeq \left\{
        \begin{gathered}
        -2(1-\omega_{0}^{2}), \\
        -2\left[1 + c_1 c_2 + \frac{2 \omega_0^2}{\omega_0^2 - 1}(1+c_2^2)\right] \frac{\omega_d^2}{2}
        +2 i \omega_0 \omega_d (c_2 - c_1).
        \end{gathered}
    \right.
\end{equation}
If $\Lambda$ has a positive real part then $\Phi$ will be linearly unstable. This yields the Eckhaus \textit{instability} criterion,
\begin{equation} \label{Eqn: Eckhaus Criterion}
    1 + c_1 c_2 + \frac{2 \omega_0^2}{\omega_0^2 - 1}(1+c_2^2) < 0. 
\end{equation}
Notice that for $\omega_0 = 0$ this reduces to \cref{Eqn: BF Criterion}. Also notice that since in general $0 < \omega_0^2 < 1$, the third term on the left-hand side is always negative. Hence, if the $\omega_0 = 0$ mode is unstable, then all modes are unstable. Conversely, if the $\omega_0 = 0$ mode is stable, then there will always be some critical value $|\omega_0| > \omega_c$ above which the Eckhaus instability will always occur, since $\omega_0^2 -1$ diverges at $\omega_0 =\pm 1$. We calculate this critical frequency in the next section. 

Last, the form of equation \eqref{Eqn: Perturbed SM} somewhat obscures the frequency of the whistler corresponding to the perturbed single-mode solution $\tilde \Phi$. By Taylor expanding the exponential in equation \eqref{Eqn: Perturbed SM}, and keeping terms to first order in $a(\zeta,\tau)$ and $\theta(\zeta,\tau)$, one has, 
\begin{equation}
    \tilde \Phi \simeq \Phi + \frac{1}{2} \sqrt{1- \omega_0^2} 
    (\tilde a_0 + i \tilde \theta_0) e^{\Lambda \zeta} 
    \left ( 
    e^{i [ k_0 \zeta - \tau(\omega_0 - \omega_d)]} +
    e^{i [ k_0 \zeta - \tau(\omega_0 + \omega_d)]}
    \right)
\end{equation}
By extracting the whistler frequency using the operator identity mentioned above, one finds that the perturbed expression corresponds to a whistler with the frequency, 
\begin{equation} \label{Eqn: Freq of Perturbed Phi}
    \omega = \omega_m + \frac{\omega_0}{t_0} \pm \frac{\omega_d}{t_0}.
\end{equation}

\subsection{Reduced Stability Criteria \& Stability Bandwidth} \label{Sec: Stability Bandwidth}

In this section, we simplify the instability criteria then use the result to show that there is no Benjamin-Feir instability and calculate the range of frequencies included in the Eckhaus stability band. Notice both instability criteria (equations \eqref{Eqn: BF Criterion} and \eqref{Eqn: Eckhaus Criterion}) depend solely on the constants $c_1$ and $c_2$, 
\begin{align}
    c_1 &\equiv - \left. \frac{\alpha_r}{\alpha_i} \right|_{\omega_s} \equiv \frac{\lambda_r''(\omega_s)}{\lambda_i''(\omega_s)} \\
    c_2 & \equiv - \left.\frac{\beta_r}{\beta_i} \right|_{\omega_s},
\end{align}
which must be evaluated at some reference frequency $\omega_s$ near the maximum of the linear growth rate $\omega_m$. Ideally, one would express the instability criteria in terms of the input parameters of the system ($v_{\perp 0}$, $\gamma_0$, etc.), rather than the more compact but less physical coefficients $c_1$ and $c_2$. However, $c_1$ and $c_2$ depend on $\lambda_0$, which is the solution to a cubic equation, so that the exact expressions for $\lambda_0$ and $\lambda_0''$ are unwieldy. 

One can simplify the analysis by taking the reference frequency $\omega_s$ to be the gyroresonance frequency, i.e. the frequency at which $\delta = 0$. Since this resonance occurs near a local maxima of $\lambda_0$, this satisfies the assumption that $\omega_s$ is near $\omega_m$. We denote the $\delta=0$ frequency by $\omega_\delta$. Now the expression for $\lambda_0(\omega_\delta)$ can be readily obtained by setting $\delta=0$ in equation \eqref{Eqn: Cubic Equation}, which yields, 
\begin{equation}
    \lambda_0(\omega_\delta) = \left. (-u g h_c)^{1/3} e^{-2\pi/3} \right|_{\omega_\delta}.
\end{equation}
Equation \eqref{Eqn: lambda appx} is exact at $\delta=0$, but the behavior at nearby points is not accurate enough to properly determine its second derivative, $\alpha$. However, by a perturbation expansion of the cubic equation, one can obtain an order-by-order approximation for $\lambda_0(\omega)$ in powers of $\delta$ which is valid near $\omega_\delta$. By assumption, $\delta$ is small for resonant interactions, so the expansion can be used to obtain an excellent approximation for $\alpha$. The perturbation assumption $\lambda_0 = f_0+f_1\delta+f_2\delta+\cdots \ $ in equation \eqref{Eqn: Cubic Equation} yields, 
\begin{equation}
    (f_0^3+ u g h_c)+(f_0^2 + 3 f_0^2f_1)\delta + (2 f_0 f_1 + 3 f_0 f_1^2 + 3 f_0^2 f_2) \delta^2 + \dots =0.
\end{equation}
Solving order by order in $\delta$, one obtains the first three terms, 
\begin{equation} \label{Eqn: lambda appx}
    \lambda_0 \simeq (-u g h_c)^{1/3} - \frac{1}{3} \delta + \frac{1}{9}(- u g h_c)^{-1/3} \delta^2.
\end{equation}
Therefore, 
\begin{equation}
\begin{aligned}
    \frac{\partial^2 \lambda_0}{\partial \omega^2}
    &\simeq
    \frac{\partial^2 (-u g h_c)^{1/3}}{\partial \omega^2}
    -\frac{1}{3}\frac{\partial^2 \delta}{\partial \omega^2} \\
    & + \frac{1}{9}\bigg[
    \frac{\partial^2 (-u g h_c)^{-1/3}}{\partial\omega} \delta^2 +
    4 \frac{\partial (-u g h_c)^{-1/3}}{\partial \omega} \frac{\partial \delta}{\partial \omega} \delta
    \\
    & + 2  (-u g h_c)^{-1/3} \left( \frac{\partial \delta}{\partial \omega} \right)^2
    + 2 (- u g h_c)^{-1/3} \delta \frac{\partial^2 \delta}{\partial \omega ^2}
    \bigg].
\end{aligned}
\end{equation}
By evaluating at $\omega_\delta$, one has,  
\begin{equation}
    \alpha \simeq 
    \frac{\partial^2 (-u g h_c)^{1/3}}{\partial \omega^2}
    -\frac{1}{3}\frac{\partial^2 \delta}{\partial \omega^2} 
    + \frac{2}{9}(-u g h_c)^{-1/3} \left( \frac{\partial \delta}{\partial \omega} \right)^2.
\end{equation}
Finally, since $\delta$ may be written as, $\delta = \Omega_{e0}/\gamma_0 - \omega +k v_{z0}$, one obtains, 
\begin{equation} \label{Eqn: alpha appx}
    \alpha \simeq \frac{2}{9} (-u g h_c)^{-1/3} \left(\frac{v_{z0}}{v_g}-1 \right)^2,
\end{equation}
where $v_g >0$, and the subdominant contributions originating from the first two terms in \eqref{Eqn: lambda appx} have been dropped (these terms are insufficient to produce parabolic behavior in $\lambda_0$ near $\omega_\delta$, thus do not contribute meaningfully to the second derivative). Importantly, we always take the root of $\lambda_0$ with the highest linear growth rate, i.e. the largest negative imaginary part, which corresponds to the branches $(-u g h_c)^{1/3} = |u g h_c|^{1/3}e^{-2 \pi i /3}$ and $(-u g h_c)^{-1/3} = |u g h_c|^{-1/3}e^{2 \pi i /3}$ in the above expressions. 

From equation \eqref{Eqn: alpha appx} one can readily show that, independent of the initial conditions, 
\begin{equation} \label{Eqn: c1 appx}
    c_1 \simeq \frac{1}{\sqrt{3}}.
\end{equation}
Using equation \eqref{Eqn: lambda appx} in equation \eqref{Eqn: beta} one has,
\begin{equation}
    \beta = 2 \sqrt{3}\frac{ u h_c}{g} e^{-i\pi/6}. 
\end{equation}
Thus $c_2$ may be written as, 
\begin{equation} \label{Eqn: c2 appx}
    c_2 = \sqrt{3}.
\end{equation}

From these expressions for $c_1$ and $c_2$, it is evident that the Benjamin-Feir instability criterion (\cref{Eqn: BF Criterion}) cannot be met, hence the Benjamin-Feir instability cannot occur, irrespective of the system parameters. This means the mode with the maximum linear growth rate is always stable and surrounded by a band of stability. 

Furthermore, the Eckhaus instability criterion (\cref{Eqn: Eckhaus Criterion}) becomes $\omega_0^2>1/5$, so that the critical onset frequency is $\omega_c = 1/\sqrt5$. Thus, the region of stability is $-\omega_c<\omega_0<\omega_c$, or approximately $-.45<\omega_0<.45$. This corresponds to a stability bandwidth for the whistler of $\Delta\omega \simeq 2\omega_c/t_0$. For typical magnetospheric parameters \cite{bonham_whistler_2025,soto-chavez_chorus_2012}, this is approximately $\Delta \omega \simeq 0.02 \, \Omega_{e0}$. This quantity refers to the bandwidth of potentially stable single modes, rather than the bandwidth of a wave packet. Although the physical interpretation differs, the value is of the same order as CLUSTER spacecraft observations of chorus bandwidths, which are typically $0.03 \,\Omega_{e0} $ to $ 0.2 \, \Omega_{e0}$ \cite{santolik_frequencies_2008}. 

\section{Numerical Simulations}

We now demonstrate the behavior of single-mode solutions within and outside of the Eckhaus stability band derived above. We solve the GLE numerically using Mathematica's standard numerical differential equation tools, employing the method of lines option to convert the PDE to an initial value problem in the continuous variable $\zeta$ by discretizing the variable $\tau$ on a grid. This results in an ODE in $\zeta$ at each grid point of $\tau$, which we integrate using the explicit modified midpoint method. Commensurate with our periodic boundary conditions in time, the $\tau$ derivatives are computed pseudospectrally. To incorporate noise while preserving the periodic boundary conditions, we add 256 harmonics of a reference frequency, $\omega_0$, with random amplitudes drawn from a uniform distribution between 0 and $10^{-3}(1-\omega_0^2)^{1/2}$. In all cases, we take $c_1=1/\sqrt3$ and $c_2=\sqrt3$, focusing on the magnetospheric conditions associated with the FEL model. 

We present three characteristic behaviors. First, single modes within the Eckhaus stability band persist under large amplitude perturbations. Second, single modes outside the Eckhaus stability band may propagate temporarily, but eventually evolve towards a stable single mode. The transition process can occur within a few wavelengths, or can occur in a more distributed manner, presenting as sweeping crescents in the spacetime diagrams. Last, an initially noisy spectrum of modes can condense to a single-mode state. 

\begin{figure}[ht] 
    \centering
    \includegraphics[width=.9\textwidth]{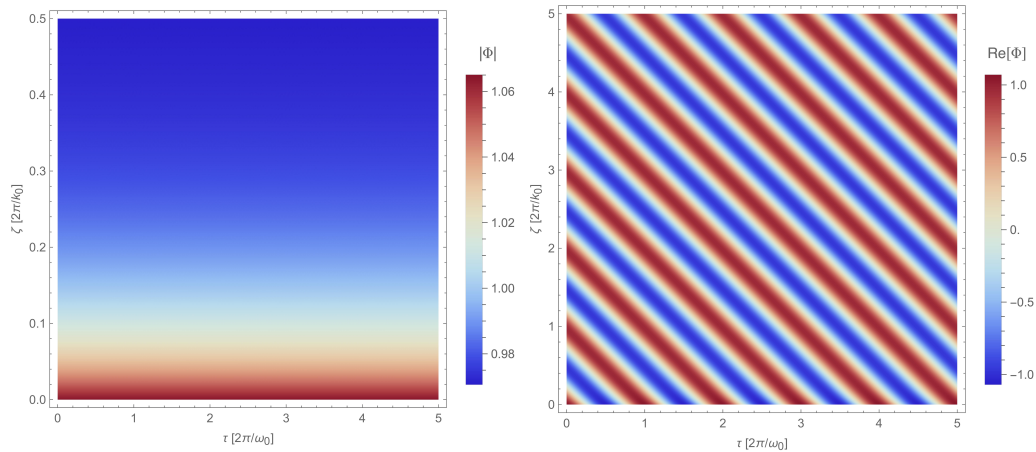}
    \caption{Spacetime plots of the amplitude (left) and real part (right) of $\Phi$, demonstrating that a stable mode ($\omega_0=1/4$) dissipates a $10\%$ amplitude perturbation within one wavelength.} 
    \label{Fig 1: Stable w_0}
\end{figure}

We begin by demonstrating the behavior of single-mode solutions within the Eckhaus stability band, choosing $\omega_0=1/4$ as a representative mode and including a moderate amplitude perturbation. Hence, the initial condition $\Phi(\zeta=0) = (1+\tfrac{1}{10})(1-\omega_0^2)^{1/2}\exp{(-i\omega_0\tau)}$. In the absence of the amplitude perturbation, this mode is an exact solution of the GLE, and will persist unchanged. Figure \ref{Fig 1: Stable w_0} shows that such steady state propagation occurs even when the amplitude is initially perturbed. As can be seen in the left panel, the amplitude perturbation decays within a fraction of a wavelength to $|\Phi|=0.97$, or the amplitude which is self-consistent with the single mode frequency-amplitude relation, $|\Phi|=(1-\omega_0^2)^{1/2}$. Thereafter, although this it is not shown in the figure, the mode persists indefinitely. 

\begin{figure}[ht] 
    \centering
    \includegraphics[width=.9\textwidth]{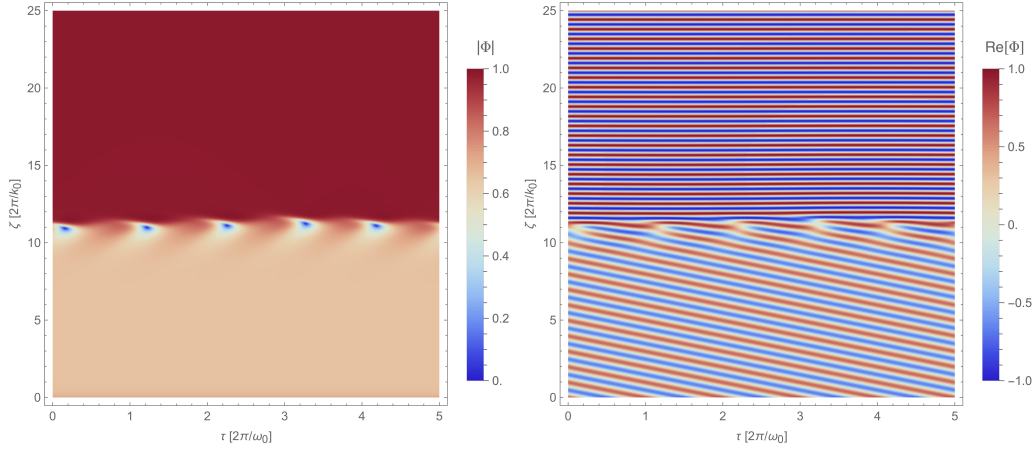}
    \caption{An unstable mode ($\omega_0=3/4$) dissipates an initial $10\%$ amplitude perturbation within one wavelength. The resulting mode, still unstable, persists for several wavelengths, then transitions to a stable mode.} 
    \label{Fig 2: Unstable w_0}
\end{figure}

Next, we consider the behavior of modes outside the stability band. Figure \ref{Fig 2: Unstable w_0} presents spacetime plots for the same initial conditions as in figure \ref{Fig 1: Stable w_0}, but with the unstable frequency $\omega_0=3/4$. So that the eventual transition to stability is not driven merely by numerical errors, in addition to the initial amplitude perturbation we include small overall constant. Hence the initial condition is, $\Phi(\zeta=0) = (1+\tfrac{1}{10})(1-\omega_0^2)^{1/2}\exp{(-i\omega_0\tau)} + 10^{-9}$. As with the previous case, the amplitude perturbation decays quickly, in a fraction of a wavelength, such that the amplitude becomes self-consistent with the frequency, $|\Phi| = (1-\omega_0^2)^{1/2} = 0.66$. The resultant mode, although unstable, is still a solution to the GLE, consistent with the fact that the mode persists for several wavelengths before defects and phase ripples appear which quickly force the system towards a stable mode. 

\begin{figure}[ht] \label{Fig 3: Unstable w_0 + Noise}
    \centering
    \includegraphics[width=.9\textwidth]{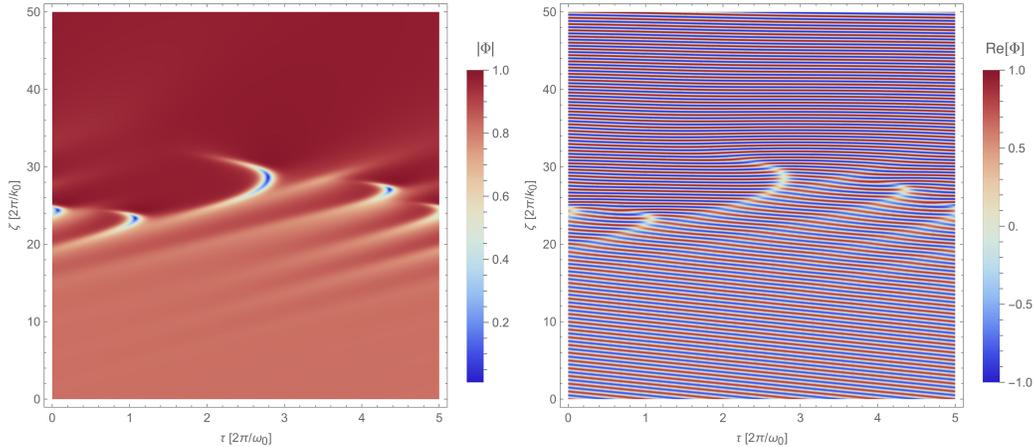}
    \caption{An unstable mode ($\omega_0=0.55$) nearer to the stability threshold gradually transitions to a stable mode due to the presence of noise.} 
\end{figure}

Having demonstrated the rapid dissipation of amplitude perturbations and the tendency of unstable modes to evolve towards stable modes, we now consider the effects of noise on a mode closer to the stability threshold. In figure \ref{Fig 3: Unstable w_0 + Noise} we consider the initial condition $\Phi(\zeta=0) = (1-\omega_0^2)^{1/2}\exp{(-i\omega_0\tau)} + \delta\Phi$, where $\omega_0=0.55$, and $\delta\Phi$ includes harmonics of $\omega_0$ with random amplitudes drawn uniformly from $|\delta\Phi| \in \left[0,(1-\omega_0)^2/10^5\right]$. In the absence of the noise, since the mode is an exact solution of the GLE, one would need to rely on numerical errors to stimulate the onset of the stability transition. Here this onset is due to the deliberate noise, so the irregularities appear within a few wavelengths of the origin. However, since this mode is closer to the stability threshold than that of figure \ref{Fig 2: Unstable w_0}, consistent with equation \eqref{Eqn: lambda appx}. the instabilities which drive the system toward a stable state have a slower growth rate. Rather than occurring over a single wavelength, the stability transition is distributed across tens of wavelengths, presenting as sweeping crescents in the spacetime diagrams. 

\begin{figure}[ht]
    \centering
    \includegraphics[width=.9\textwidth]{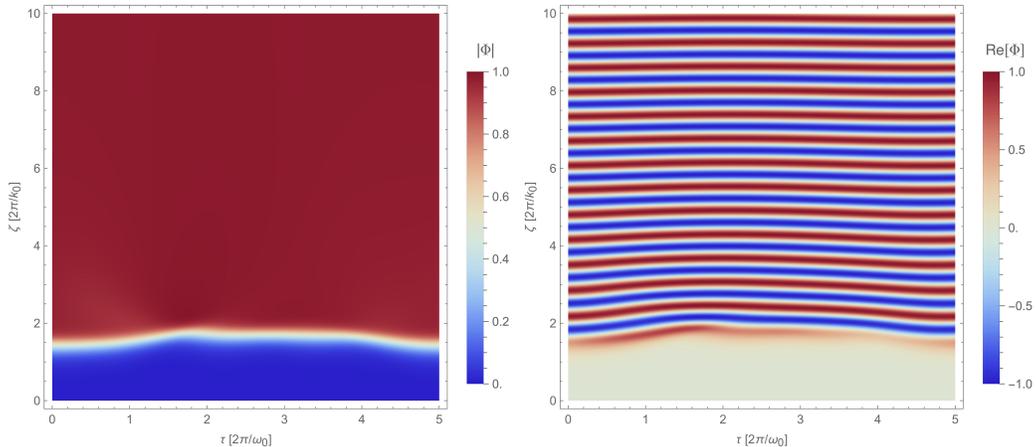}
    \caption{Condensation of 256 modes with small random amplitudes down to a single stable mode. }
    \label{Fig 4: Mode Condensation}
\end{figure}

Finally, in figure \ref{Fig 4: Mode Condensation} we consider an initial condition consisting solely of a noisy spectrum of modes with small random amplitudes. The noise consists of harmonics of the reference frequency $\omega_0=3/4$ with random amplitudes drawn uniformly from the range $\delta\Phi_m\in\left[0,(1-\omega_0^2)^{1/2}/ 10^{3}\right]$, so that the initial condition is, $\Phi = \sum_{m=-128}^{128}|\delta\Phi_m| e^{-im\omega_0\tau/N_\tau}$, where $N_\tau=5$ is the number of periods included in the range. The broad spectrum of modes present in the initial state quickly transfer their energy to the mode with the highest linear growth rate. This behavior, known as mode condensation, has been investigated in the context of FELs as a potential mechanism for obtaining beam purity \cite{ng_ginzburg-landau_1998}. 

\section{Discussion \& Conclusion}

In this study, we investigated the stability of single-mode chorus propagation in the magnetosphere using the Ginzburg-Landau equation. This equation, recently proposed as a nonlinear description for whistler-mode chorus, originates from the FEL model \cite{bonham_whistler_2025, soto-chavez_chorus_2012}. By drawing upon insights from the FEL literature, we found that the single mode with the highest linear growth is stable and surrounded by a narrow band of stability, outside of which modes are vulnerable to the Eckhaus instability. After briefly reviewing the linear stability analysis, by a perturbation expansion of the linear growth rate, we found that the width of the Eckhaus stability band is approximately two percent of the background electron cyclotron frequency. We then conducted a brief numerical study to illustrate three key behaviors - the persistence of stable frequencies under amplitude perturbations, the evolution of unstable frequencies towards stable ones, and the condensation of a spectrum of random modes down to a single mode. 

Finally, we note an additional implication of this work of interest to the theory of chorus. From equation \eqref{Eqn: SM Solution} it can be readily seen that the single-mode solutions to the GLE contain an explicit self-consistent coupling between the amplitude and frequency of the mode, namely $|\Phi| =(1-\omega_0^2)^{1/2}$. Thus the amplitude evolution that occurs for Eckhaus-unstable modes, as seen in figures \ref{Fig 2: Unstable w_0} and \ref{Fig 3: Unstable w_0 + Noise}, implies frequency evolution. In fact, distinct frequencies can be seen in the right panels of those figures, where the contours of the real part of the mode undergo a change in angle before and after the stability transition. Because the initial value of $\omega_0$ may be above or below the stability band, the frequency evolution may be downwards (if $\omega_0>\omega_c$) or upwards (if $\omega_0<-\omega_c$) as the mode transitions into the stable band. The implications for magnetospheric chorus, including the frequency chirp rate and duration, will be the subject of a forthcoming publication. 

%%%%%%%%%%%%%%%
%% Chapter 4 %%
%%%%%%%%%%%%%%%

\chapter{Conclusions} \label{Chapter 4: Conclusions}

\section{Limitations \& Future Research}

The FEL model of magnetospheric chorus has shown promise as a description of whistler-electron interactions near the magnetic equator. The novel contributions presented in this work - e.g. the Stuart-Landau equation for the wave evolution and the prediction of solitary whistlers - highlight its advantages. However, the model is still in its beginning stages. The following is a brief list of limitations and open questions. Each is, of course, also a potentially fruitful direction for further research. 

Magnetospheric chorus is generally characterized not only by its discrete narrow-band structure but also by a tendency for these discrete modes to sweep upwards and/or downwards in frequency - a phenomenon known as chirping. Hence, a strong theory of chorus should also describe these frequency dynamics. It is generally accepted that there are two main mechanisms for frequency chirping, the first being nonlinear amplitude evolution and the second being inhomogeneities in the background magnetic field. While the former is a prominent feature in the current model, the latter is not. However, the original FEL model \cite{soto-chavez_chorus_2012} included an extra term in the equations of motion to account for background field inhomogeneity. Therefore, with some effort, in principle both chirping mechanisms can be explored from within the context of the FEL model. 

A priori, it is surprising that a coupled interaction described by 2$N$+2 equations can be reduced to three first-order nonlinear differential equations. In this sense, the collective variable approach has again been shown to be of great utility. However, the question of the regime of validity of the collective variable equations presented here has been left unexplored. Notably, equations \crefrange{Eqn: NLCVE 1}{Eqn: NLCVE 3} demonstrate strong agreement, over several post-saturation oscillation cycles, between the collective variable equations and the 2$N$+2 equations they approximate. Slippage in the amplitude oscillation cycles, however, can be seen. In addition, after a few cycles the collective variable amplitude exhibits a dip that is not present in the 2$N$+2 equations. Typically, small deviations in nonlinear systems lead to large deviations over time. Here, however, since at saturation the electrons are trapped and oscillating in a potential well, both solutions are quasi-periodic, and qualitative agreement seems to persist over arbitrarily long timescales. We note that it may be possible to improve the approximate constant of motion $H_0$ (\cref{Eqn: Cons 2}) to an exact one, and thereby achieve reduced amplitude slippage, but given that the agreement is already very strong, it is not clear that any additional insight would be gained into the system. 

Thus far the model has focused primarily on the wave behavior. Along those lines, one outstanding question is the stability profile of the solitary wave solutions to the GLE. These have been noted here as valid solutions to the GLE and their approximate widths and amplitudes have been derived. However, whether or not they are likely to be found in nature also depends on their stability profile. A rigorous stability analysis such as that conducted for the single-mode solutions would shed light on this matter. In contrast to the detailed analysis of the wave properties predicted by the FEL model, the model's implications for the electron dynamics have been left mostly unexplored. A related limitation is its use of a monoenergetic electron beam in the initial state, which simplifies many of the calculations. While this assumption deviates from a more realistic distribution, a preliminary simulation of \crefrange{Eqn: EOM 1}{Eqn: EOM 3} with a Maxwellian distribution indicates that the electron bunching and subsequent amplitude evolution still occurs similarly in either case. This is somewhat expected, since even when simulating the 2$N$+2 equations with an initially monoenergetic beam, the system immediately develops a velocity spread. It is also likely that the introduction of additional spatial degrees of freedom would introduce new effects in the system. 

\section{Summary}

This dissertation presents recent developments of the free-electron laser (FEL) model of whistler-mode chorus in the magnetosphere. In the first chapter, we provide an introduction to whistler-mode chorus, including a derivation of the cold plasma whistler dispersion relation. We also introduce the concept of a FEL, and derive the foundational equations for the FEL model of whistler-mode chorus (\crefrange{Eqn: EOM 1}{Eqn: EOM 3}). These equations are cast in the form of FEL equations, and juxtaposed with well-known FEL equations from the literature. Due to the multiplicity of parameters appearing throughout the work, a glossary of symbols is provided in \autoref{Glossary of Symbols}. 

In the second chapter, we reduce this large set of 2$N$+2 coupled nonlinear differential equations (where $N$ is the number of resonant electrons) down to just three coupled nonlinear differential equations using the method of collective variables. Using this result, we study the amplitude behavior of the whistler in the nonlinear phase of the interaction. We show that when the nonlinear terms are dropped, we recover the linear collective variable equations of \citet{soto-chavez_chorus_2012}. By a perturbation approach, we combine the nonlinear collective variable equations into a Stuart-Landau equation (SLE) (\cref{Eqn: SLE}) for the whistler amplitude and phase which includes a linear growth term and a first-order nonlinearity. This simple equation, involving just two constants, can be used to estimate the saturation amplitude and timescale. By extending the SLE to accommodate a whistler wave packet containing a spectrum of modes with spatially dependent amplitudes, we derived a Ginzburg-Landau equation (GLE) (\cref{Eqn: GLE Full,Eqn: GLE Std.}) for the multi-mode behavior of chorus in space and time. 

Among the interesting multi-frequency phenomena predicted by the GLE is the existence of solitary waves - a nonlinear phenomenon characterized by the propagation of a wave packet without variation in its shape or speed. Solitary waves are analogous to solitons, but lack their collisional robustness. From the GLE presented in this work, we predict the properties of solitary whistler waves in the Van Allen radiation belts. In addition to various multi-frequency behaviors, the GLE also contains single-mode solutions, which are the subject of the third chapter. After first presenting the single-mode solutions to the GLE, we conduct a linear stability analysis by adding small oscillatory perturbations to them, reproducing instability conditions well known in the GLE and FEL literature. By a perturbation expansion of the linear growth rate, we reduce these inequalities to simple forms which reveal that the system is only linearly stable within a narrow frequency band. Outside of this band, which surrounds the mode with the highest linear growth rate, the Eckhaus instability occurs. We then conduct numerical simulations to support these results, demonstrating the robustness of the stable frequencies and the evolution of the unstable frequencies into the stability band. Last, we present numerical evidence for the existence and tendency towards mode condensation, whereby a single stable mode evolves out of an initially noisy spectrum of modes with small random amplitudes. 

%%%%%%%%%%%%%%%%
%% Appendices %%
%%%%%%%%%%%%%%%%

\begin{appendices} 

%%%%%%%%%%%%%%%%%%%%%%%%%
%% Glossary of Symbols %%
%%%%%%%%%%%%%%%%%%%%%%%%%

\chapter{Glossary of Symbols} \label{Glossary of Symbols}

\textit{With the exception of the symbols appearing exclusively in the derivation of the dispersion relation (\cref{Sec: Dispersion Relation}), which is both introductory and notationally self-contained, this glossary contains each mathematical symbol used within the manuscript, organized conceptually. Some items, belonging naturally to more than one category, can be found in more than one subsection. Each symbol is accompanied by a brief description and/or mathematical definition. The symbol $\zeta$ is used in two different senses, which may be discerned by the context. In the subsection entitled Accoutrements (notational accessories), the capital letter $O$ serves as a placeholder for a symbol acted on by some operator, e.g. $\bar O \equiv O/v_g$.}
\\

% Glossary Table Headings

\begin{longtable}{@{}p{0.19\textwidth}p{0.79\textwidth}@{}}
\textbf{Symbol} & \textbf{Definition / Description} \\
\midrule
\endfirsthead
\endhead

% Glossary Entries

\multicolumn{2}{@{}l}{\textbf{Fundamental Constants}}\\
$e$ & Magnitude of charge of electron. \\
$m_e$ & Mass of electron. \\
$\epsilon_0$ & Permittivity of free space. \\
$\mu_0$ & Permeability of free space. \\
$c$ & Speed of light in vacuum. \\
\addlinespace

\multicolumn{2}{@{}l}{\textbf{Accoutrements}}\\
$\bar O$ & Division by $v_g$, $\bar O \equiv O/v_g$. \\
$\langle O\rangle$ & Average over resonant particles, $ \langle O\rangle\equiv \frac{1}{N}\sum_{j=1}^N O_j$. \\
$O_r$, $O_i$ & Real and imaginary parts of $O$ (except in the case of $K$). \\
$|O|$ & Magnitude of the complex number $O$. \\
$O^*$ & Complex conjugate of $O$. \\
$\dot O$ & Time derivative of $O$. \\
$O$ + c.c. & $O$ plus its complex conjugate, $O + \text{c.c.} \equiv O+O^*$ \\
\addlinespace

\multicolumn{2}{@{}l}{\textbf{Geomagnetic Field}}\\
$\bm B_0$ & Background magnetic field at geomagnetic equator. \\
$L$ & McIlwain $L$-shell parameter. \\
$z$ & $z$ coordinate, positive along $\bm B_0$. \\
\addlinespace

\multicolumn{2}{@{}l}{\textbf{Whistler Wave}}\\
$\bm B_w$ & $\bm B_w \equiv B_w(t) (\cos \varphi, \sin \varphi,0) $ \\
$B_w$,$B_w(t)$ & Amplitude of $\bm B_w$ \\
$\varphi$ & Phase of $\bm B_w$; $\varphi \equiv \omega(k)t - k z + \phi(t)$ \\
$\phi, \phi(t)$ & Slowly varying phase of whistler.  \\
$\omega$, $\omega(k)$ & Whistler frequency. \\
$k$ & Whistler Wavenumber. \\
$v_g$ & Whistler group velocity, $v_g \equiv \partial\omega/\partial k = 2c\frac{\omega^{1/2} (\Omega_{e0} - \omega)^{3/2}}{\Omega_{e0}\omega_{pe}} \approx 2 s\Omega_{e0}/k$. \\
$s$ & Dispersion factor, $s \equiv \frac{\omega}{\Omega_{e0}-\omega}=\frac{c^2k^2}{\omega_{pe}^2}$.\\
\addlinespace

\multicolumn{2}{@{}l}{\textbf{Collective Variables}}\\
$A$ & Collective variable / scaled complex whistler field, $A\equiv b\,e^{-i\psi_0}$. \\
$\varphi_A$ & Complex phase of $A$, i.e. $A = |A|e^{i\varphi_A}$. \\
$X$ & Collective variable / bunching parameter, $X\equiv \langle e^{i\Delta\psi}\rangle$. \\
$X_0$ & Initial value of $X$. \\
$Y$ & Collective variable / $\sim$ phase-weighted momentum change, $Y \equiv\langle \Delta\eta_z\,e^{i\Delta\psi}\rangle$. \\
$b$ & Scaled complex whistler field, $b=\frac{e}{m_e} B_w\,e^{i\phi}$. \\
$\psi$ & Shifted whistler–electron phase difference, $\psi \equiv \theta-\omega t+k z = \theta - \varphi + \phi $. \\
$\psi_j$ & $\psi$ for $j$\textsuperscript{th} resonant electron. \\
$\psi_0$ & $d\psi_0/dt \equiv (d\psi/dt)|_{t=0} \equiv \delta \Rightarrow \psi_0 = \delta t + \text{const.}$  \\
$\Delta\psi$ & $\Delta \psi \equiv \psi-\psi_0$. \\
$\Delta\psi_j$ & $\Delta\psi$ for $j$\textsuperscript{th} resonant electron. \\
\addlinespace

\multicolumn{2}{@{}l}{\textbf{Resonant Electrons}}\\
$N$ & Number of resonant electrons. \\
$n_r$ & Number density of resonant electrons. \\ 
$\theta$ & Electron gyrophase angle with respect to $x$-axis. \\
$ v_\perp$,$v_{\perp0}$ & Speed perpendicular to $\bm B_0$; assumed constant throughout. \\
$v_{z0}$ & Initial speed parallel to $\bm B_0$. \\ 
$v_r$ & Whistler-electron gyro-resonance velocity, $v_r \equiv (\omega-\Omega_{e0}/\gamma_0)/k$. \\
$\bm\eta$ & Proper velocity, $\bm \eta=\gamma \bm v$. \\ 
$\eta_z$ & Parallel proper velocity, z component of $\bm \eta$. \\
$\eta_\perp$ & Perpendicular proper velocity, component of $\bm \eta$ perpendicular to $\bm B_0$. \\
$\eta_{z0}$ & Initial value of $\eta_z$; equivalent for all resonant electrons. \\
$\Delta \eta_z$ & Change in parallel proper velocity, $\Delta \eta_z \equiv \eta_z - \eta_{z0}$. \\
$\Delta \eta_z^j$ & $\Delta \eta_z$ for $j$\textsuperscript{th} resonant electron. \\
$\Delta \eta_z^2$ & Square of $\Delta\eta_z$, $\Delta\eta_z^2 = (\eta_z - \eta_{z0})^2$. \\ 
$\gamma$ & Lorentz factor, $\gamma \equiv (1-\bm v^2/c^2)^{-1/2} \equiv (1+\bm\eta^2/c^2)^{1/2} = \gamma_{\perp0}(1+\eta_z^2/c^2)^{1/2}$. \\
$\gamma_0$ & Initial value of $\gamma$. \\
$\gamma_j$ & $\gamma$ for $j$\textsuperscript{th} resonant electron. \\
$\gamma_{\perp0}$ & Perpendicular Lorentz factor, $\gamma_{\perp0} \equiv (1 - v_{\perp0}^2/c^2)^{-1/2}$. \\
\addlinespace

\multicolumn{2}{@{}l}{\textbf{Collective Variable Equations}}\\
$g$ & Current coupling constant, $g \equiv \frac{s \Omega_{e0} \omega_{pr}^2 u}{k c^2}$. \\
$\delta$ & Detuning constant, $\delta = k (v_{z0}-v_r)$. \\
$h_c$ & $h_c \equiv -(\Gamma_{-1} \Omega_{e0}+k/\gamma_0 +\Gamma_{-1}k\eta_{z0}) = -\frac{k}{\gamma_0}(1- \frac{\gamma_{\perp 0}^2\eta_{z0}^2}{\gamma_0^2 c^2} -\frac{\gamma_{\perp 0}^2\Omega_{e0}\eta_{z0}}{\gamma_0^2 c^2 k})$. \\
$s$ & Dispersion factor, $s \equiv \frac{\omega}{\Omega_{e0}-\omega}=\frac{c^2k^2}{\omega_{pe}^2}$.\\
$u$ & Half perpendicular speed, $u\equiv\frac{v_{\perp0}}{2}$. \\
$\Gamma_n$ & Initial value of $n$\textsuperscript{th} derivative ($n>0$) of $\gamma$, $\Gamma_n  \equiv \frac{d^n\gamma}{d\eta_z^n}|_{\eta_{z0}}$. \\
$\Gamma_{-n}$ & Initial value of $n$\textsuperscript{th} derivative ($n>0$) of $1/\gamma$, $\Gamma_{-n}  \equiv \frac{d^n\gamma^{-1}}{d\eta_z^n}|_{\eta_{z0}}$. \\
$P_0$ & Momentum-like conserved quantity, $P_0=\langle\Delta\eta_z\rangle+\frac{u}{g}|A|^2$. \\
$H_0$ & Hamiltonian-like conserved quantity, $ H_0=\frac{\langle\Delta\eta_z^2\rangle}{2}-\frac{u}{h_c}(A^*X+X^*A)+\frac{u\delta}{g h_c}|A|^2$. \\
\addlinespace

\multicolumn{2}{@{}l}{\textbf{Ginzburg-Landau \& Stuart-Landau Equations}}\\
$\lambda_0$, $\lambda_0(\omega)$ & Dominant linear growth rate of $B_w$, $\lambda_0^3+\delta\lambda_0^2+u g h_c \equiv 0$. \\
$\beta$ & Cubic nonlinearity coefficient, $\beta \equiv -2 u h_c \lambda_0 (  -\frac{1}{g\lambda_0} + \frac{ \delta}{g \lambda_0^2} 
    + \frac{u h_c}{\lambda_0^4} + \frac{u h_c}{|\lambda_0|^4})$. \\
$\mu$ & Defined by, $\mu\equiv1+ \left. \frac{\partial\lambda_0}{\partial\omega}\right|_{\omega_s}$. \\
$\alpha$ & Dispersion of growth rate, $\alpha \equiv \left.\frac{\partial^2\lambda_0}{\partial\omega^2}\right|_{\omega_s}$. \\
$c_1$ & Defined by, $c_1 \equiv -\alpha_r/\alpha_i$. \\
$c_2$ & Defined by, $c_2 \equiv -\beta_r/\beta_i$. \\
\addlinespace

\multicolumn{2}{@{}l}{\textbf{Casting Ginzburg-Landau Equation in Standard Form}}\\
$\zeta$ & Scaled space, $\zeta=z/z_0$. \\
$\tau$ & Scaled time, $\displaystyle \tau=\frac{t}{t_0}+\frac{z}{v_0 t_0}$. \\
$z_0$ & Length scale in mapping, $\displaystyle z_0=\left(-\bar\lambda_i+\frac{\bar\mu_i^{\,2}}{2\bar\alpha_i}\right)^{-1}$. \\
$t_0$ & Time scale in mapping, $\displaystyle t_0=\sqrt{\frac{\bar\alpha_i z_0}{2}}$. \\
$v_0$ & Speed scale in mapping, $\displaystyle v_0=\left(-\bar\mu_r+\frac{\bar\alpha_r\,\bar\mu_i}{\bar\alpha_i}\right)^{-1}$. \\
$\Phi_0$ & Amplitude scale in mapping, $\Phi_0 = \sqrt{\frac{1}{z_0 \bar\beta_i}}$.\\ 
$K_0$ & $\zeta$ coefficient in mapping, $\displaystyle K_0=z_0\!\left(\bar\lambda_r-\frac{\bar\alpha_r\,\bar\mu_i^{\,2}}{2\bar\alpha_i^{\,2}}\right)$. \\
$\Omega_0$ & $\tau$ coefficient in mapping, $\displaystyle \Omega_0=\frac{\bar\mu_i\, t_0}{\bar\alpha_i}$. \\
\addlinespace

\multicolumn{2}{@{}l}{\textbf{Frequencies and Rates}}\\
$\Omega_{e0}$ & Electron cyclotron frequency due to $\bm B_0$, $\Omega_{e0} \equiv e B_0/m_e$. \\
$\omega_{pr}$ & Resonant-electron plasma frequency, $\omega_{pr} \equiv \sqrt{\frac{n_{r}e^2}{\epsilon_0 m_e}}$. \\
$\omega_{pe}$ & Background electron plasma frequency, $\omega_{pe} \equiv \sqrt{\frac{n e^2}{\epsilon_0 m_e}}$. \\
$\omega_s$ & Reference frequency about which $\lambda(\omega)$ is expanded in GLE derivation. \\
$\omega_m$ & Frequency at which $\lambda_0(\omega)$ is maximized (near $\delta$ =0). \\
$\omega_d$ & Frequency of perturbations to single-mode solution to GLE. \\
$\omega_c$ & Positive value of $\omega_0$ at which Eckhaus instability occurs. \\
$\delta\omega$ & Frequency bandwidth estimate from width of $\lambda_0(\omega)$, $\delta\omega \approx 2\sqrt{-2\lambda_i/\alpha_i}$. \\
$\Delta \omega$ & Bandwidth of Eckhaus stable single-mode solutions. \\
\addlinespace

\multicolumn{2}{@{}l}{\textbf{Solitary Waves}}\\
$Q$ & Solitary wave amplitude factor, $|Q| = 2 \left( 1 + \frac{c_1\sigma-1}{\sigma^2 + 2 c_1 \sigma - 1}\right)^{1/2}$. \\
$\Omega$ & Solitary wave phase factor, $\Omega \equiv c_1 - \frac{2 \sigma(1+c_1)}{\sigma^2 +2 c_1 \sigma -1}. $\\
$\sigma$ & Solitary wave exponent parameter, $\displaystyle \sigma^2-3\frac{c_1c_2+1}{c_1-c_2}\sigma-2=0$. \\
$K$ & Solitary wave width parameter, $K=(\sigma^2+2c_1\sigma-1)^{-1/2}$. \\
$K_r$,$K_i$ & Real and imaginary solutions (not parts) of $K$. \\ 
$\Delta t$ & Pulse width of solitary waves, $\Delta t \equiv 2 t_0/|K|$. \\
$\Delta T$ & Pulse separation between periodic solitary waves, $\Delta T \equiv \pi t_0/|K|$. \\
\addlinespace

\multicolumn{2}{@{}l}{\textbf{Single-Mode Solutions of GLE}}\\
$\Phi$ & Single-mode solution, $\Phi = \sqrt{1-\omega_0^2}\exp[ i(k_0\zeta-\omega_0\tau+\phi_0)]$. \\
$\omega_0$ & Frequency of single-mode, $\omega_0 \in (-1,1)$. \\
$k_0$ & Wavenumber of single-mode, $k_0 \equiv -c_2-(c_1-c_2)\omega_0^2$. \\
$\phi_0$ & Phase of single-mode at $\zeta=\tau=0$, assumed zero throughout. \\
\addlinespace

\multicolumn{2}{@{}l}{\textbf{Single-Mode Stability Analysis}}\\

$\tilde\Phi$ & Perturbed single-mode. \\
$\delta \Phi$ & Perturbations to $\Phi$ in numerical solutions. \\
$\omega_d$ & Frequency of small perturbations to single-mode. \\
$a(\zeta,\tau)$ & Small perturbation to amplitude of single-mode, $a(\zeta,\tau) \equiv \tilde a(\zeta) \cos(\omega_d \tau)$. \\
$\tilde a(\zeta)$ & Spatial part of $a(\zeta,\tau)$. \\
$\tilde a_0$ & Origin value of $\tilde a (\zeta)$, $\tilde a_0 \equiv \tilde a(\zeta=0)$. \\
$\theta(\zeta,\tau)$ & Small perturbation to phase of single-mode, $\theta(\zeta,\tau) \equiv \tilde \theta(\zeta) \cos(\omega_d \tau)$ \\
$\tilde \theta(\zeta)$ & Spatial part of $\theta(\zeta,\tau)$. \\
$\tilde \theta_0$ & Origin value of $\tilde \theta(\zeta)$,  $\tilde \theta_0 \equiv \tilde \theta(\zeta=0)$. \\
$\Lambda$ & Growth rate of perturbations, $\tilde \theta = \tilde \theta_0\exp(\Lambda \zeta)$ and $\tilde a = \tilde a_0\exp(\Lambda \zeta)$. \\
$\omega_c$ & Positive value of $\omega_0$ at which Eckhaus instability occurs. \\
$f_1,f_2,f_3,\dots$ & Coefficients of $\delta$ in expansion of $\lambda_0$ about $\delta=0$. \\
$\Delta \omega$ & Bandwidth of Eckhaus stable single-mode solutions. \\

\end{longtable}

\end{appendices}

%%%%%%%%%%%%%%%%
%% References %%
%%%%%%%%%%%%%%%%

\bibliography{references}

%%%%%%%%%
%% End %%
%%%%%%%%%

\end{document}